\documentclass[aps,prd,twocolumn,showpacs,preprintnumbers,nofootinbib]{revtex4}
 
 \usepackage{graphicx}
 \usepackage[FIGTOPCAP]{subfigure}
 \usepackage{dcolumn}
 \usepackage{bm}
 \usepackage{amsmath}
 \usepackage{epstopdf}
 \usepackage{amsmath,amssymb}
 \usepackage{bigints}
 \usepackage{mathtools}
 \usepackage{array}
 \usepackage{bbm}

 \newcommand{\beq}{\begin{eqnarray}}
 \newcommand{\eeq}{\end{eqnarray}}

 \newcommand{\real}{{\sf I}\kern-.12em{\sf R}}
 \newcommand{\comp}{{\sf I}\kern-.50em{\sf C}}
 \newcommand{\unity}{{\sf I}\kern-.54em{\sf 1}}

 \newcommand{\tr}{\mbox{Tr}}

\begin{document}
 
 \title{
 Spectrum of the Laplace-Beltrami Operator and the Phase Structure of Causal Dynamical Triangulation
 }
 
 \author{Giuseppe Clemente}\email{giuseppe.clemente@pi.infn.it}
 \author{Massimo D'Elia}\email{massimo.delia@unipi.it}
 \affiliation{Dipartimento di Fisica dell'Universit\`a di Pisa and INFN
 	- Sezione di Pisa,\\ Largo Pontecorvo 3, I-56127 Pisa, Italy.}

 \date{\today}
 
 \begin{abstract}
 We propose a new method to characterize the different phases 
 observed in the non-perturbative 
numerical approach to quantum gravity known as
 Causal Dynamical Triangulation.
 The method is based on the analysis of the eigenvalues and the eigenvectors
 of the Laplace-Beltrami operator computed on the triangulations: 
 it generalizes previous works based on the analysis of diffusive processes and
proves capable of providing 
more detailed information on the geometric properties of the triangulations.
 In particular, we apply the method to the analysis of spatial
 slices, showing that the different phases can be characterized
 by a new order parameter related to
the presence or absence of a gap in the spectrum
 of the Laplace-Beltrami operator, 
and deriving an effective dimensionality
 of the slices at the different scales.
 We also propose quantities derived from the spectrum that could 
 be used to monitor the running to the continuum limit around
 a suitable critical point in the phase diagram, if any is found.
 \end{abstract}
 
 
 \maketitle

 \section{Introduction}
 
Causal Dynamical Triangulations (CDT)~\cite{cdt_report} is a numerical Monte-Carlo approach to Quantum Gravity based on the Regge formalism, 
where the path-integral is performed over geometries represented by simplicial manifolds called ``triangulations''. The action employed is a discretized version of the Einstein-Hilbert one, and the causal condition of global hyperbolicity is enforced on triangulations by means of a space-time foliation.
 
One of the main goals of CDT is to find a critical 
point in the phase diagram where the continuum limit can be performed in the form of a second-order phase transition. The phase diagram shows the presence of four different phases~\cite{cdt_secondfirst,Ambjorn:2014mra,Ambjorn:2015qja,cdt_charnewphase,cdt_newhightrans,cdt_toroidal_phasediag}, and the hope is that the transition lines
separating some of these phases could contain such a second order critical point.
Presently, such phases are identified by order parameters 
which are typically based on the counting of
the total number of simplexes of given types or on other similar quantities (e.g., the coordination number of the vertices
of the triangulation).
The main motivation of the present study is to enlarge the set 
of observables available for CDT, 
trying in particular to find new
order parameters and to better characterize 
the geometrical properties of the various phases 
at different scales.

One successful attempt to characterize the geometries of CDT has been obtained by implementing diffusion
processes on the triangulations~\cite{cdt_spectdim,Coumbe:2014noa}. In practice, one analyzes
the behavior of {\em random walkers} moving around the triangulations:
from their properties (e.g., the return probability) one can derive
relevant information, such as the effective dimension felt at different
stages of the diffusion (hence at different length scales). In this way, estimates of 
the {\em spectral dimension} of the triangulations have been obtained.

In this paper we propose and investigate a novel set of observables for 
CDT configurations, based on spectral methods, namely, the analysis of the 
properties of the eigenvalues and the eigenvectors of the 
Laplace--Beltrami (LB) operator. 
This can be viewed as a generalization of the analysis of the spectral 
dimension, since the Laplace--Beltrami operator completely specifies the 
behavior of diffusion processes 
(see Appendix~\ref{sec:heatkernel} for a closer comparison).
Still, as we will show in the following, 
the Laplace--Beltrami operator contains more geometric 
information than just the spectral dimension.

Nowadays, spectral methods find application in a huge variety 
of different fields. To remember just a few of them, we mention 
shape analysis in computer aided design and medical physics~\cite{reuter_dna,reuter_cad}, dimensionality reduction and spectral clustering for feature selection/extraction in machine learning~\cite{eigenmaps}, optimal ordering in the PageRank algorithm of the Google Search engine~\cite{pagerank}, connectivity and robustness analysis of random networks~\cite{robustnetw}. Therefore, the 
application to CDT is just one more application of a well known 
analysis tool. On the other hand, some well known results which have
been established in other fields will turn out to be useful in our
investigation of CDT.

In the present paper, we limit our study to the LB spectrum of spatial
slices. Among the various results, we will show that 
the different phases can be characterized
 by the presence or absence of a gap in the spectrum
 of the LB operator, as it happens
for the spectrum of the Dirac operator
in strong interactions, and we will give an interpretation
of this fact in terms of the geometrical properties 
of the slices. The presence/absence of a gap
will also serve to better characterize the 
two different classes of spatial slices which are found 
in the recently discovered bifurcation phase~\cite{Ambjorn:2014mra,Ambjorn:2015qja,cdt_charnewphase,cdt_newhightrans}.
Moreover, we will show how
the spectrum can be used to derive 
an effective dimensionality of the triangulations 
at different length scales, and to investigate quantities useful 
 to characterize the critical behavior
expected around a possible second order transition point.

The paper is organized as follows.
In Section~\ref{sec:ctdreview} we discuss our numerical setup together
with a short review of the CDT approach,
summarizing in particular the major features of the phase diagram 
that will be useful for the discussion of our results. 
In Section~\ref{sec:LBproperties} we describe some of the most relevant 
properties of the Laplace-Beltrami operator in general, then focusing 
on its implementation for the spatial slices of CDT configurations
and discussing a toy model where the relation between
the LB spectrum and the effective dimensionality of the system
emerges more clearly. Numerical results are discussed in 
Section~~\ref{sec:numres_eigenvalues}. 
Finally, in Section~\ref{sec:conclusions}, we draw our conclusions 
and discuss future perspectives. 
Appendix~\ref{sec:heatkernel} is devoted to a discussion of 
the relation existing between the spectrum of the LB operator
and the spectral dimension, defined by diffusion processes 
as in Ref.~\cite{cdt_spectdim}.
 
 \section{A brief review on CDT and numerical setup}\label{sec:ctdreview}
 
   It is well known that, perturbatively, 
General Relativity without matter is 
non-renormalizable already at the two-loop level~\cite{sagnotti}. 
Nevertheless, interpreted in the framework of the 
Wilsonian renormalization group approach~\cite{wilsonian_RG}, this really 
means that the gaussian point in the space of parameters of the theory is 
not an UV fixed point, as for example it happens for asymptotically 
free theories. 
Indeed, Weinberg conjecture of \emph{asymptotic safety} of the 
gravitational interaction~\cite{weinbergsASS} states the existence of an UV non-gaussian fixed point, which makes the theory well defined in the UV 
(i.e.~renormalizable), but in a region of the phase diagram not accessible
 by perturbation theory. Various non-perturbative methods have been developed in the 
last decades to investigate this possibility, 
like Functional Renormalization Group 
techniques~\cite{qeg}
or the Monte-Carlo simulations of standard Euclidean 
Dynamical Triangulations (DT)~\cite{edt1,edt2,dt_forcrand,dt_syracuse} or 
Causal Dynamical Triangulations, the latter being the subject of 
this study.

Monte-Carlo simulations of quantum field theories are based on the 
path--integral formulation in Euclidean space, where 
expectation values 
of any observable $\mathcal{O}$ 
are estimated as averages over field configurations sampled with probability proportional to $e^{-\frac{S}{\hbar}}$, $S$ being the action functional of the theory. 
 Regarding the Einstein--Hilbert theory of gravity, the action is a functional of the metric field $g_{\mu \nu}$, given by\footnote{For simplicity, we are not including manifolds with boundaries, so there is no Gibbons--Hawking--York term in the action.}
 \begin{equation}\label{eq:act_cont}
 S[g_{\mu\nu}] = \frac{1}{16 \pi G} \bigintssss d^dx \,\sqrt{-g}\,(R -2 \Lambda)
 \end{equation}
 where $G$ and $\Lambda$ are respectively the \emph{Newton} and \emph{Cosmological} constants, while the path--integral expectation values are formally written as averages over geometries (classes of diffeomorphically equivalent metrics)
 \begin{equation}\label{eq:cdtove_EH_PI}
 \langle \mathcal{O} \rangle = \frac{1}{Z} \bigintssss \mathcal{D}[g_{\mu\nu}]\, \mathcal{O}[g_{\mu\nu}]e^{-\frac{S[g_{\mu\nu}]}{\hbar}\,},
 \end{equation}
 where $Z$ is the partition function.

 The first step in setting up Monte-Carlo simulations is the choice of a specific regularization of the dynamical variables into play. In the case of gravity without matter fields the only variable is the geometry itself, which can be conveniently regularized in terms of \emph{triangulations}, namely a collection of \emph{simplexes}, elementary building blocks of flat 
spacetime, glued together to form a space homeomorphic to a topological manifold. The simplexes representing (spacetime) volumes in $4$--dimensional spaces are called \emph{pentachorons}, analogous to \textit{tetrahedra} in $3$--dimensional spaces and \textit{triangles} in $2$--dimensional spaces (i.e.~surfaces).
 
 Besides the general definition, and at variance with standard DT, triangulations employed in CDT simulations are required to satisfy also a causality condition of global hyperbolicity\footnote{The global hyperbolicity condition is equivalent to the existence of a Cauchy surface, the strongest causality condition which can be imposed on a manifold~\cite{causconds}.}. This is realized by assigning an integer time label to each vertex of the triangulation in order to partition them into distinct sets of constant time called \emph{spatial slices}, and constraining simplexes to fill the spacetime between adjacent slices (i.e.~slices with neighbouring integer labels). The resulting triangulation has therefore a \emph{foliated} structure\footnote{The main reason for restricting to foliated triangulations is that it allows to define conveniently the analytical continuation from Lorentzian to Euclidean space (see Ref.~\cite{cdt_report} for details). However, simulations without preferred foliation in $2+1$ dimensions have been build in Ref.~\cite{cdt_nofoliae}, showing results similar to the foliated case.}, and the simplexes can be classified by a (time--ordered) pair specifying the number of vertices on the slices involved (e.g., the pairs $(4,1)$, $(3,2)$, $(2,3)$ and $(1,4)$ classify all spacetime pentachorons).
 In order to ensure both the simplicial manifold property and the foliated structure \textit{at the same time}, spatial slices, considered as simplicial submanifolds composed of glued spatial tetrahedra, 
need to be topologically equivalent. This basically means that triangulations are always geodetically complete manifolds, and topological obstructions (e.g., singularities) can only be realized in an approximate fashion, with increasing accuracy in the thermodynamic limit (infinite number of simplexes). The numerical results shown in Section~\ref{sec:numres_eigenvalues} refer to slices with $S^3$ topology, but other topologies could be investigated as well (e.g., the toroidal one~\cite{cdt_toroidal,cdt_toroidal_phasediag}).
 
 In practice, it is convenient, without loss of generality, to impose a further condition, that is fixing the length--squared of every spacelike link (i.e.~connecting vertices on the same slice) to a constant value $a^2$, and the square--length of every timelike link (i.e.~connecting vertices on adjacent slices) to a constant value $-\alpha a^2$. The constant $a$ takes the role of \emph{lattice spacing}, while $\alpha$ represents a genuinely regularization--dependent asymmetry in the choice of time and space discretizations. With this prescription, simplexes in the same class (according to the above definition) not only are equivalent topologically, but also geometrically, so that the expression of the discretized action greatly simplifies. Indeed, at the end of the day\footnote{A series of steps is needed in order to obtain the CDT action in Eq.~\eqref{eq:4Daction}: Regge discretization of the continuous action, computation of volumes and dihedral angles, Wick rotation and the use of topological relations between simplex types.}, the standard $4$--dimensional action employed in CDT simulations with $S^3$ topology of the slices and periodic time conditions becomes a functional of the 
triangulation $\mathcal{T}$, and takes the relatively simple form
 \begin{equation}\label{eq:4Daction}
 S_E = -k_0 N_0 + k_4 N_4 + \Delta ( N_4 + N_{41}-6 N_0) \, ,
 \end{equation}
 where $N_0$ counts the total number of vertices, $N_4$ counts the total number of pentachorons, and $N_{41}$ is the sum of the total numbers of type $(4,1)$ and type $(1,4)$ pentachorons, while $k_4$, $k_0$ and $\Delta$ are free dimensionless parameters, related to the Cosmological constant, the Newton constant, and the freedom in the choice of the time/space asymmetry parameter $\alpha$ (see Ref.~\cite{cdt_report} for more details).
 
 We want to stress that, even if CDT configurations are defined by means of triangulations, the ultimate goal of the approach is to perform a continuum limit in order to obtain results describing continuum physics of quantum gravity. Therefore,
the specific discretization used in CDT must be meant as artificial, becoming irrelevant in the continuum limit. For this reason, simplexes should not be considered as forming the physical fabric of spacetime: eventually, 
one would like to find a critical point in the parameter 
space where the correlation length diverges and the memory about
the details of the fine structure is completely lost.
\\

 In standard CDT simulations, configurations are sampled using a Metropolis--Hastings algorithm~\cite{metropolishastings}, where local modifications of the triangulation at a given simulation time (i.e.~insertions or removals of simplexes) are accepted or rejected according to the probability induced by the action~in Eq.~\eqref{eq:4Daction} and complying with the constraints discussed above. 
 
 Unlike usual lattice simulations of quantum field theories, the total spacetime volume of CDT triangulations changes after a Monte Carlo update. In order to take advantage of finite size scaling methods (i.e.~extrapolation of results to the infinite volume limit), it is convenient to control the volume by performing a Legendre transformation from the parameter triple $(k_4,k_0,\Delta)$ to the triple $(V,k_0,\Delta)$, where the parameter $k_4$ is traded for a target volume $V$. In practice, this is implemented by a fine tuning of the parameter $k_4$ to a value that makes the total spacetime or spatial volumes\footnote{The total spatial volume of a configuration is the number of spatial tetrahedra, which equals $\frac{N_{41}}{2}$ by elementary geometrical arguments.} fluctuate with mean around a chosen target volume (respectively $\overline{N}_{4}$ or $\overline{N}_{41}$), and adding to the sample only configurations whose total volume lies in a narrow range around the target one.
 Moreover, a (weak) spacetime volume fixing to a target value $\overline{N}_{4}$ can be enforced, for example, by adding a term to the action of the form $\Delta S = \epsilon (N_{4}-\overline{N}_{4})^2$,
 where $\epsilon$ quantifies how much large volume fluctuations are suppressed. A relation similar to the latter holds for fixing the total spatial volume (substituting $N_{41}$ with $N_{4}$). 
 Fixing a target total spatial volume $V_{S,tot}=\frac{\overline{N}_{41}}{2}$, one can investigate the properties of configurations sampled at different values of the remaining free parameters $k_0$ and $\Delta$. 

\begin{figure}
 	\centering
 	\includegraphics[width=1.0\linewidth]{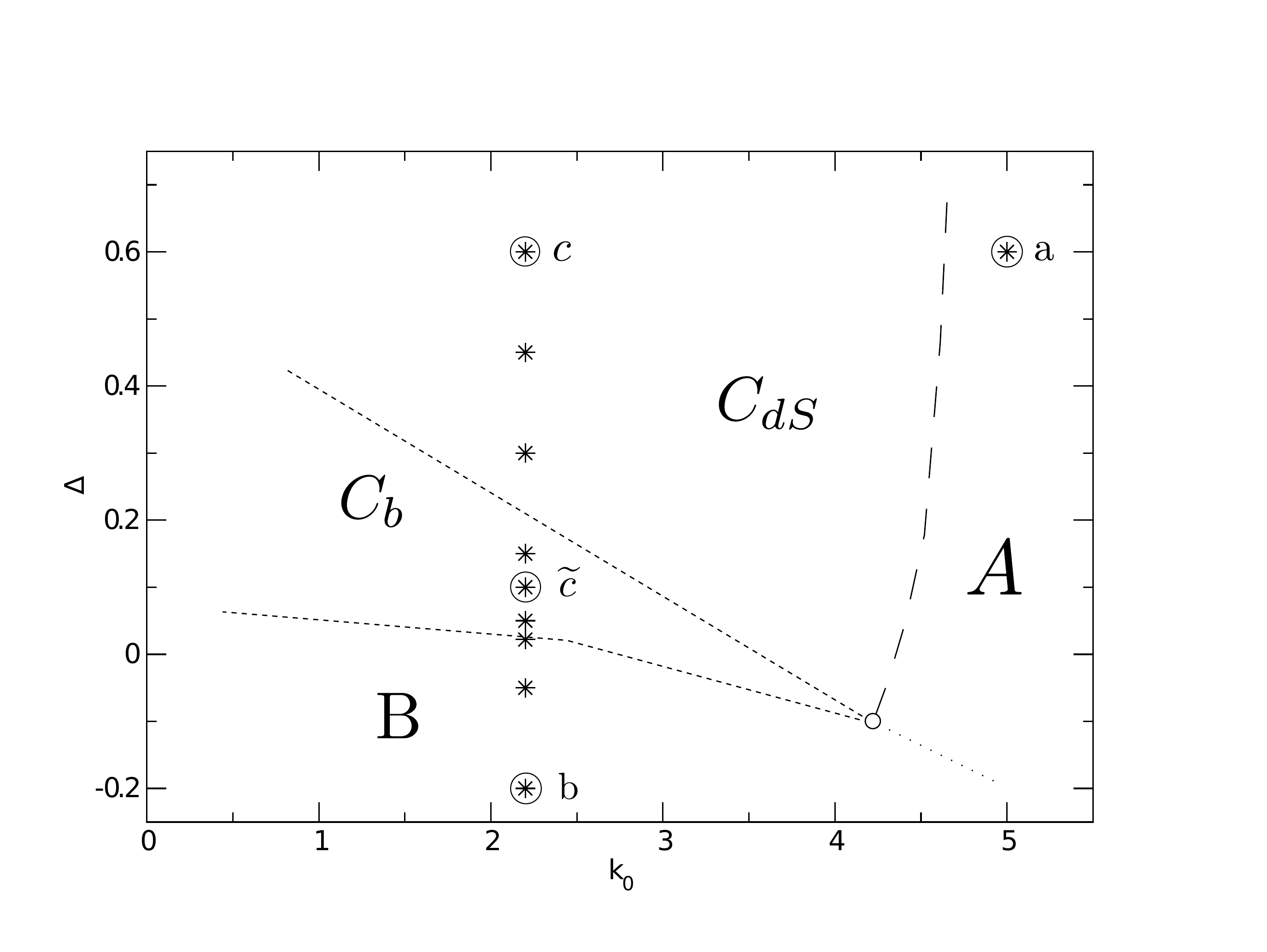}
 	\caption{Sketch of the phase diagram CDT in $4$d and with spherical topology of spatial slices. The results shown in the present paper have been obtained from simulations running at the points marked by a star symbol $*$. The circled and labeled points $a$,$b$,$c$ and $\widetilde{c}$ refer to simulations running deeply inside the respective phases (see Table~\ref{tab:simpoints}). The position of transition lines is only qualitative.}
 	\label{fig:phasediag}
 \end{figure}

 The general phase structure of CDT which is found in the 
$k_0$ - $\Delta$ plane is thoroughly discussed in the literature~\cite{cdt_report,cdt_secondord,cdt_newhightrans,cdt_charnewphase}. Here we will only recall some useful facts.
Four different phases have been identified, 
called $A$, $B$, $C_{dS}$ and $C_{b}$, as sketched 
in Fig.~\ref{fig:phasediag}, where for the two $C$ phase the labels $dS$ and $b$ stand
respectively for {\em de Sitter} and {\em bifurcation}.
 At a qualitative level, configurations in the different phases can be 
characterized by the distribution of their spatial volume $V_S(t)$, 
which counts the number of spatial tetrahedra (spatial volume $V_S$) in each 
slice as a function of the slice time $t$. For configurations in the $B$ phase, 
the spatial volume is concentrated almost in a single slice, leaving the other slices 
with a minimal volume\footnote{Triangulations with $S^3$ topology, like spatial slices in $3+1$ standard 
CDT simulations, must have at least 5 tetrahedra.}. For both the $C_{dS}$ and $C_{b}$ phase, the 
spatial volume is peaked at some slice--time but then, unlike the case of the $B$ phase, falls off more gently 
with $t$, so that the majority of the total spatial volume is localized in a so-called ``blob'' with a finite 
time extension; also in this case, slices out of this blob have a minimal volume. 
Finally, configurations in phase $A$ are 
characterized by multiple and uncorrelated peaks in the spatial volume distribution.

 From these observations, it is apparent that $C_{dS}$ and $C_{b}$ are the only physically relevant phases. Indeed, 
the average spatial volume distribution in the $C_{dS}$ phase is in good agreement with 
the prediction for a de Sitter Universe, having a $S^4$ geometry after analytical continuation to the 
Euclidean 
space~\cite{cdt_desitter}. The bifurcation phase, instead, is characterized by the presence of two different
classes of slices which alternate each other in the slice time $t$~\cite{cdt_charnewphase,cdt_newhightrans}.

 The transition lines between the different phases (dashed lines in Fig.~\ref{fig:phasediag}) have been investigated 
by means of convenient observables.
Regarding the $B$-$C_b$  and $A$-$C_{dS}$ transition lines, the definitions employed 
are based on the observation that changes in the qualitative behavior of the spatial volume distribution function 
$V_S(t)$ occur for almost constant values of $\Delta$ or $k_0$ respectively, suggesting quantities conjugated to 
them in the action \eqref{eq:4Daction} as candidate order parameters: 
namely $conj(\Delta)\equiv (N_4+N_{41}-6 N_0)/N_4$ for the $B$-$C_b$ 
transition and $conj(k_0)\equiv N_0/N_{41}$ for the $A$-$C_{dS}$ transition. Finite size scaling computations 
using these observables as order parameters suggest a first-order nature for the $A$-$C_{dS}$ transition, 
while the $B$-$C_b$ transition appears to be of 
second-order~\cite{cdt_secondord}. The definition of observables employed as order parameters for the 
$C_{b}$-$C_{dS}$ transition is more involved~\cite{cdt_charnewphase,cdt_newhightrans}: in the $C_b$ phase, one of the two classes of spatial slices is characterized by the presence of vertices 
with very high coordination number; also in this case there are hints for a 
second-order transition, even if results might depend
on the topology chosen for the spatial slices~\cite{cdt_toroidal_phasediag}.
 	
 \begin{table}
 	\centering
 	\begin{tabular}{c l r r }
 		&	$k_0$	& $\Delta$	&  phase	\\ \hline
 		$b$		&	2.2		&	-0.2	&	$B$		\\ \hline	
 		&	2.2		&	-0.05	&	$B$		\\ 
 		&	2.2		&	0.022	&	$B$	\\ 
 		&	2.2		&	0.05	&	$C_b$	\\ \hline			
 		$\widetilde{c}$	&	2.2		&	0.1		&	$C_b$	\\ \hline		
 		&	2.2		&	0.15	&	$C_b$	\\ 
 		&	2.2		&	0.3		&	$C_{dS}$\\ 
 		&	2.2		&	0.45	&	$C_{dS}$\\ \hline			
 		$c$		&	2.2		&	0.6		&	$C_{dS}$\\ \hline	
 		$a$		& 	5		&	0.6		&	$A$		\\ 
 	\end{tabular}
 	\caption{Coordinates ($k_0$ and $\Delta$) for the simulation points chosen, as shown in Figure~\ref{fig:phasediag}, and the phases in which they are contained. Some of the points are labeled also by a letter for later convenience. The assignment of simulation points to the different phases refers 
to the total volumes fixed in our runs ($N_{41} = 40k$ and $80k$).}\label{tab:simpoints}
 \end{table}

Global counts of simplexes, like those entering the definitions of $conj(\Delta)$ and $conj(k_0)$, 
are not sufficient to clearly distinguish the different geometrical properties of the various 
phases. From this point of view, the \emph{spectral dimension} $D_S(\tau)$ 
(see Appendix~\ref{sec:heatkernel} for more details) is probably one of the few useful probes
available up to now to probe the geometrical structure of CDT configurations. 
It is basically a measure of the effective dimension of the geometry at different stages of the diffusion process,
it has permitted to demonstrate that, in the bulk of configurations in the de Sitter phase, 
the spectral dimension tends to a value $D_S\simeq 4$ for large diffusion times~\cite{cdt_spectdim}.
In the following, we will show how the analysis of the spectrum of the LB operator, which is 
discussed in the following section, permits to access
new classes of observables, and how some clear characteristic differences among the various phases 
emerge in this way.
\\

  The code employed for this study is an home--made implementation in C++ of the standard CDT algorithm discussed 
in Ref.~\cite{cdt_report}, which was checked against
many of the standard results which can be found in the CDT literature.
  We performed simulations with parameters chosen as shown in Fig.~\ref{fig:phasediag}
by points marked with a star symbol and reported also in Table~\ref{tab:simpoints};
for later convenience, four points, each being deep into one of the 4 phases, 
have been labeled by a letter: $a$, $b$, $c$ and $\widetilde{c}$.
For most simulation points we have performed simulations with two different 
total spatial volumes, $V_{S,tot}=20k$ and $V_{S,tot}=40k$, 
adopting a volume fixing parameter $\epsilon = 0.005$;
we have verified that our results are independent 
of the actual prescription used.

 \section{The Laplace-Beltrami operator}\label{sec:LBproperties}
 
 The LB operator, usually denoted by the symbol $-\Delta$, is the generalization of the standard Laplace operator. Its specific definition depends on the underlying space and on the algebra of functions on which it acts.
 For a generic smooth Riemannian  manifold $(\mathcal{M},g_{\mu \nu})$ the Laplace-Beltrami operator acts on the algebra of smooth functions $f\in \mathcal{C}^\infty(M)$ in the form~\cite{diffgeom}:
 \begin{eqnarray}
\label{eq:LB-Mg}
 -\Delta f &=& -\frac{1}{\sqrt{|g|}} \partial_\mu (\sqrt{|g|} g^{\mu \nu} \partial_\nu f) \\
&=& - g^{\mu \nu} (\partial_\mu \partial_\nu -\Gamma_{\mu \nu}^\alpha \partial_\alpha) f \, , \nonumber
 \end{eqnarray}
 where $g$ is the metric determinant, $g^{\mu \nu}$ is the inverse metric and $\Gamma_{\mu \nu}^\alpha$ are the Christoffel symbols.
 
 It is easily shown that $-\Delta$ is invariant with respect to isometries. Furthermore, since it is positive semi-definite, a set of eigenvectors $\mathcal{B}_{M}$ solving the eigenvalue problem $-\Delta f = \lambda f$ is an orthogonal basis for the algebra $\mathcal{C}^\infty(M,\mathbb{R})$; in the following we will refer to such sets as \textit{spectral bases}, which, for convenience and without loss of generality, we will always consider orthonormal.
 A spectral basis can then be used to define the Fourier transform as basis change from real to momentum space (e.g. sines and cosines in $\mathbb{R}^n$, or spherical harmonics in $S^2$), while the eigenvalues associated to each eigenspace contain information about the characteristic scales of the manifold. 

We will now elaborate further on the interpretation of the spectrum of eigenvalues, considering a diffusion process on a generic manifold $M$ described by the heat equation
 \begin{align}\label{eq:Mdiffeq}
 \partial_t u(x,x_0;t) - \Delta u(x,x_0;t) = 0 \, .
 \end{align}
 We can expand the solution in a spectral basis $\mathcal{B}_M = \big\{ e_{n} | \lambda_n \in \sigma_M, 
\lambda_{n+1} \geq \lambda_n    \big\}$ associated to the spectrum of (increasingly ordered) eigenvalues $\sigma_M = \big\{ \lambda_n \big\}$ 
 \begin{align}
 u(x;t) = \sum_{n=0}^{|\sigma_M|-1} u_{n}(t) e_{n}(x) \, ,
 \end{align}
 so that Eq.~\eqref{eq:Mdiffeq} is transformed (by orthogonality) in a set of \emph{decoupled} equations
 \begin{align}
 \partial_t u_{n}(t) = -\lambda_n u_{n}(t)\;\, \forall\, n\, ,
 \end{align}
 \begin{align}\label{eq:decmodes}
 \implies u(x;t) = \sum_{n=0}^{|\sigma_M|-1} e^{-\lambda_n t} u_{n}(0)  e_{n}(x) \, .
 \end{align}
 In the form of Eq.~\eqref{eq:decmodes} the geometric role of eigenvectors in the diffusion process is evident: $\lambda_n$ represents the diffusion rate of the mode $e_{n}(x)$, so that the smallest eigenvalues are associated to eigenvectors along the slowest diffusion directions and vice versa. In this specific sense, the spectrum $\sigma_{M}$ encodes information about the characteristic scales of the manifold, while the set of eigenstates $\mathcal{B}_{M}$ identifies all the possible diffusion modes, and forms a basis for the algebra of functions on the manifold. Similar considerations can be applied to the problem of wave propagation on the manifold, where the heat equation is replaced by the wave equation; this is the reason behind the famous idea of ``hearing the shape of a drum''~\cite{drumshape_rev}.
 
 The definition of the Laplace--Beltrami operator can be extended easily to more general algebras, like the graded algebra of differential forms or the algebra of functions on a graph~\cite{spectra_of_graphs,spectral_graph_theory}, the latter being of particular importance in our discussion, since, as discussed below, it allows us to implement straightforwardly the spectral analysis on CDT spatial slices, by means of their associated dual graphs. 
 A undirected graph $G$~\cite{graph_theory} is formally a pair of sets ($V$,$E$), where $V$ contains \emph{vertices}, which assume the role of lattice sites, whereas the set of edges, $E \subset V \times V$, is a symmetric binary relation on $V$ encoding the connectivity between vertices in the form of ordered pairs of vertices $\{(v_i, v_j)\}$. 
The reason why, in this first study, we choose to apply spectral methods to analyze the geometry of spatial slices only is that spatial tetrahedra have all link lengths equal to the spatial lattice size $a$, so that the distance between their centers is equal for any adjacent tetrahedra; therefore, it is possible to represent faithfully spatial slices by dual undirected and unweighted graphs, where the vertex set is the set of tetrahedra, and the edge set is the adjacency relation between tetrahedra.
 The algebra on which the Laplace-Beltrami operator acts can be taken as that of the real-valued functions $f:V\to \mathbb{R}$, 
which can be represented as the vector space $\mathbb{R}^{N}$ (where $N=|V|$), once an ordering of the vertices $i\mapsto v_i \in V\;\forall i \in \{0,1,\dots,N-1\}$ has been arbitrarily chosen, without loss of generality\footnote{Every ordering can be obtained as a permutation of the canonical basis vectors.}. 
In this representation the Laplace--Beltrami operator becomes formally a matrix, named \emph{Laplace matrix}, and defined as:
\begin{equation}\label{eq:LBmat}
L = D - A \, ,
\end{equation}
where $D$ is the (diagonal) \emph{degree matrix} such that the element \mbox{$D_{i i} \equiv |\{e \in E |  v_i \in e \}|$} counts the number of vertices connected to the vertex $v_i$, while $A$ is the symmetric \emph{adjacency matrix} such that the element \mbox{$A_{i j} = \mathbb{\chi}_{E}(\{v_i,v_j\})$} is $1$ only if the vertices $v_i$ and $v_j$ are connected (i.e.~$\{v_i,v_j\}\in E$) and zero otherwise.
 
For instance, the graph associated with a one-dimensional hypercubic lattice with $N$ sites and periodic~boundary conditions corresponds to \mbox{$D = 2 \cdot\,\mathbbm{1}_{N\times N}$}, and  \mbox{$A_{i j} = \delta_{i,(j+1)\mod{N}} + \delta_{i,(j-1)\mod{N}}$}, while the Laplace matrix can be read off as the lowest order approximation to the Laplace--Beltrami operator estimated by evaluating functions on lattices sites:
 \begin{equation}\label{eq:lap1D_f}
 -\Delta f (x_i) = -\frac{d^2f}{dx^2}(x_i) = \frac{2 f_i-f_{i+1}-f_{i-1}}{a^2} + \mathcal{O}(a) \, ,
 \end{equation}
 where $a$ is the lattice spacing and $f_n\equiv f(x_{(n\bmod{N})})$.

 Notice that, since any tetrahedron of CDT spatial slices is adjacent to exactly $4$ neighboring tetrahedra, the dual graphs are $4$-regular (i.e.~each vertex has degree $4$), so that the adjacency matrix suffices to compute eigenvalues and eigenvectors ($L= 4 \cdot \,\mathbbm{1} - A$), and furthermore it is sparse. In practice, we build and save the graphs associated to each slice in the adjacency list representation. Being already a memory-efficient storage for the adjacency matrix of the graph, these structures can be directly fed to any numerical solver optimized for the computation of eigenvalues and eigenvectors of sparse, real and symmetric matrices. The spectra and eigenvectors analyzed in the present paper have been obtained using the `Armadillo' C++ library~\cite{armadillo} with Lapack, Arpack and SuperLU support for sparse matrix computation.
 
 By solving the eigensystem for the LB spectrum, we can easily obtain eigenvectors as a side product. Even if the spectrum of a graph does contain much geometric information, still alone it is not capable to completely characterize geometries, but only classes of isospectral graphs. Conversely, the joint combination of eigenvalues and eigenvectors yields 
complete information on the graph\footnote{Recall that, by the spectral theorem, the LB matrix can be decomposed as $L=U \Lambda U^t$, where $\Lambda$ is the diagonal matrix of eigenvalues, and $U$ is the matrix with corresponding eigenvectors as columns. The adjacency matrix, which defines the graph, can be simply obtained as minus the off-diagonal part of the LB matrix.}, but decomposed in a way useful for the analysis of geometries.
 
 \subsection{General properties of the eigenvalues of the Laplace matrix on graphs}

Here we will describe some 
results from spectral graph theory that allow us to extract the information mentioned above. 
For convenience, we will always consider the basis of eigenvectors $\mathcal{B}_G = \{\vec{e}_{n}\}$ to be real and orthonormal, since in this case the spectral theorem for real symmetric matrices applies.
 
 First of all we observe that, if no boundary is present, the Laplace matrix always has the 
zero eigenvalue, with a multiplicity equal to the number of connected components\footnote{This observation is not restricted to Laplace matrices of graphs, but applies to the spectra of the Laplace operator in a general space.}. For graphs made of a single connected component, any eigenfunction associated to the zero eigenvalue is simply a multiple of the uniform function $\vec{e}_{0} = \frac{1}{\sqrt{|V|}} \vec{1}_{|V|}$, where we indicate with $\vec{1}_{|V|}$ the vector in $\mathbb{R}^{|V|}$ with $1$ on each entry. Furthermore, the sum of the components of each eigenvector $\vec{e}_{n}$, with the exception of $\vec{e}_{0}$, is zero, since $\sum_{v\in V} e_{n}(v) = (\vec{e}_{n},\sqrt{|V|} \vec{e}_{0}) = 0$ by orthogonality of the chosen basis $\mathcal{B}_G$.
 In the following, we will only discuss properties of graphs with a single connected component\footnote{The more general case of graphs with multiple connected components can be easily treated by studying its components individually: the Laplace matrix can be put in block-diagonal form, one block for each component, so that its spectrum is the union of the individual spectra, and its eigenspaces are direct sums of the individual eigenspaces.}, like the ones occurring in CDT.
 
 \subsubsection{Spectral gap and connectivity}\label{subsec:spectralgap}
 As argued above, geometric information about the large scales comes from the smallest eigenvalues and associated eigenvectors.
 The $0$-th eigenvalue has a topological character, and in the general case its multiplicity tells us how many connected components the graph is composed of, but for connected graphs its role is trivial and uninteresting.
 
 Arguably the most interesting eigenvalue is the first (non-zero) $\lambda_1$, which, depending on the context, is called \emph{spectral gap} or \emph{algebraic connectivity}.
 The latter name comes from the observation that the larger the spectral gap $\lambda_1$, the more the graph is connected.
  
 A measure of connectivity for a compact Riemannian manifold $\mathcal{M}$ is given by the \emph{Cheeger isoperimetric constant} $h(\mathcal{M})$ defined as the minimal area of a hypersurface $\partial A$ dividing $\mathcal{M}$ into two disjoint pieces $A$ and $\mathcal{M}\setminus{A}$
 \begin{equation}
 h(\mathcal{M}) \equiv \inf \frac{vol(\partial A)}{vol(A) vol(\mathcal{M}\setminus{A})} \, ,
 \end{equation}
 where the infimum is taken over all possible connected submanifolds $A$.
 
 For a graph $G=(V,E)$, the Cheeger constant is usually defined by
 \begin{equation}\label{eq:CheegerDef}
 h(G) \equiv \min \Big\{\frac{|\partial A|}{|A|}\,|\, A\subset V, |A| \leq \frac{|V|}{2}\Big\} \, ,
 \end{equation}
 where $\partial A$ is the set of edges connecting $A$ with $V\setminus A$.
 The relation between the Cheeger constant and the spectral gap for a graph $G$ where all vertices have exactly $d$ neighbours is encoded in the \emph{Cheeger's inequalities}
 \begin{equation}\label{eq:Cheeger_ineq}
 \frac{1}{2} \lambda_1 \leq h(G) \leq \sqrt{2 d \lambda_1} \, . 
 \end{equation}
 This property of the spectral gap is interesting for the analysis of geometries of slices in CDT, since, as we will se in the next section, it highlights different behaviors for the various phases.
 
 \subsection{Eigenvalue distribution and a toy model}\label{subsec:toymodel}
 
When one considers
 the whole spectrum of the LB operator, two particularly
 interesting quantities are the density $\rho(\lambda)$, defined
 so that $\rho(\lambda)\, d \lambda$ gives the number of eigenvalues
 found in the range $[\lambda, \lambda + d \lambda]$, 
and its
 integral $n(\lambda)$, which gives the total number 
 of eigenvalues below a given value $\lambda$. 
 
 Both functions can be defined for single configurations (spatial
 slices) or can be given as average quantities over the 
 Euclidean path integral ensemble. As we shall see, the latter
 quantity, $n(\lambda)$, will prove particularly useful
 to characterize the properties of triangulations at different 
 scales. It is an increasing function of $\lambda$ and its inverse
 is simply the $n$-th eigenvalue $\lambda_n$. We will usually show 
$\lambda_n$ as a function of $n$ since, when considering a sample
of configurations, taking the average of $\lambda$ at fixed (integer) $n$ is easier.
 
 There are various well known results regarding the two quantities
 above, most of them involving the LB operator on smooth manifolds.
 In particular, Weyl law~\cite{weylslaw_1,weylslaw} gives the asymptotic (large $\lambda$) 
 behavior of $n(\lambda)$:
 \begin{equation}\label{eq:weyl_law}
 n(\lambda) = \frac{\omega_d}{(2 \pi)^d} V \lambda^{d/2} 
 \end{equation}
 where $V$ is the volume of the manifold (which is assumed to be 
 finite, with or without a boundary), $d$ is its dimensionality,
 and $\omega_d$ is the volume of the $d$-dimensional ball of unit radius.
 As we shall better discuss below, Weyl law, even if asymptotic,
 is generally expected 
 to hold with a good approximation 
 in the range of $\lambda$ for which one is not sensitive to the 
 specific infrared properties (i.e.~shape, boundaries 
 and/or topology) of the manifold. How violations to the 
 Weyl law emerge and how they can be related to a sort of effective 
 dimension at a given scale will be one of the main points
 of our discussion.

 \begin{figure}[t!]
 	\centering
 	\includegraphics[width=1.0\linewidth]{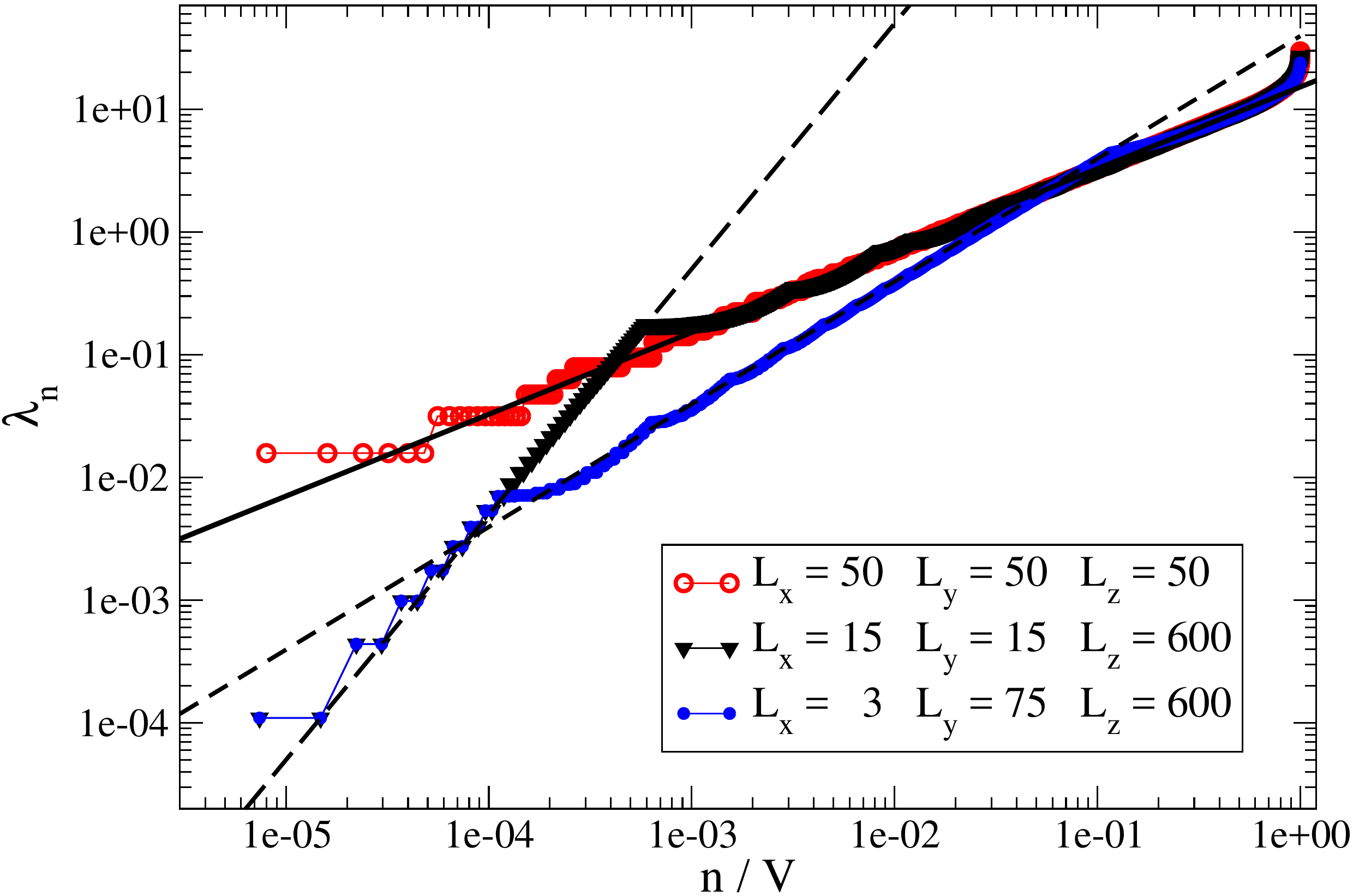}
 	\caption{Plot of $\lambda_n$ against its volume-normalized order $n/V$, for a hypercubic lattice with periodic boundary conditions (i.e., toroidal) and different combinations of sizes $L_i$ for each direction. 
 	The straight continuous line is the exact Weyl scaling, 
see Eq.~\eqref{eq:weyl_law}, predicted for $d = 3$; 
the dashed straight lines correspond to effective Weyl scalings 
for effective dimensions $d=2$ and 3.}
 	\label{fig:toymodel_1}
 \end{figure}

 In the following we shall consider the LB spectrum computed on
 discretized manifolds. It is therefore
 useful to start by analyzing a simplified and 
 familiar model, consisting of a regular and 
 finite 3-dimensional cubic lattice, with respectively 
 $L_x$, $L_y$ and $L_z$ sites along the $x$, $y$ and $z$ directions.
 All lattice sites are connected with 6 nearest neighbors sites, with
 periodic boundary conditions in all directions: this is therefore
 the discretized version of a 3-dimensional torus.
 The Laplacian operator can be simply discretized on this lattice 
 and 
 its eigenvectors coincide with the normal modes of a corresponding system
 of coupled oscillators: they are 
 plane waves
 having wave number $\vec k = (k_x, k_y, k_z)$,
 with $k_i = 2 \pi\, m_i / L_i$ and $m_i$ integers such that 
 $-L_i/2 < m_i \leq L_i/2$,
 so that
 \begin{equation}
 \lambda_{\vec m} = 4 \pi^2 \left( 
 \frac{m_x^2}{L_x^2} +
 \frac{m_y^2}{L_y^2} +
 \frac{m_z^2}{L_z^2}
 \right) \, .
 \end{equation}
 Determining $n(\bar\lambda)$ for a given $\bar\lambda$ 
 now reduces to counting how many 
 vectors $\vec m$ exist
 such that $ \lambda_{\vec m} \leq  \bar\lambda$. 
 That corresponds
 to finding the triplets of integer numbers, i.e.~the cubes
 of unit side,
 within the ellipsoid of semiaxes $R_i = \bar \lambda^{1/2} L_i / (2 \pi)$,
 with the constraint that $ - L/2 < m_i \leq L/2\ \forall\, i$.
 The latter constraint expresses the particular (cubic) discretization 
 that we have adopted for the 3-dimensional torus, i.e.~the 
 structure of the system at the UV scale: if $\bar \lambda$ is
 low enough so that $R_i < L_i\ \forall\ i$, then we are not sensitive 
 to such scale. On the other hand, the discretized structure of the 
 eigenvalues expresses the finiteness of the system, i.e.~the properties
 of the system at the IR scale: if we have also 
 $R_i \gg 1\ \forall\ i$ then we are not sensitive to such scale either,
 and the counting reduces approximately to estimate 
 the volume of the ellipsoid, so that 
 \begin{equation}
 n(\bar \lambda) \simeq 
 \frac{4 \pi}{3} R_x\,  R_y\,  R_z  =  
 \frac{4 \pi}{3} \frac{L_x\, L_y\, L_z}{(2 \pi)^3} \bar \lambda^{3/2} \, ,
 \end{equation}
 which is nothing but Weyl law for $d = 3$.
 
 In Fig.~\ref{fig:toymodel_1} we show the exact distribution of 
 $\lambda_n$ as a function of $n/V$, for various choices of 
 $L_x$, $L_y$ and $L_z$. The tick line represents the Weyl law prediction,
 $\lambda = 6 \pi^2 (n/V)^{2/3}$. 
 When  $n/V \to 1$, all systems show similar deviations
 from the law, which are related to the common structure at the 
 UV scale. The Weyl law is a very good approximation for lower 
 values of $n/V$, as expected, and actually
 down to very small values of $n/V$ for the symmetric 
 lattice where $L_x = L_y = L_z = 50$. 
 
 \begin{figure}[t!]
 	\centering
 	\includegraphics[width=1.0\linewidth]{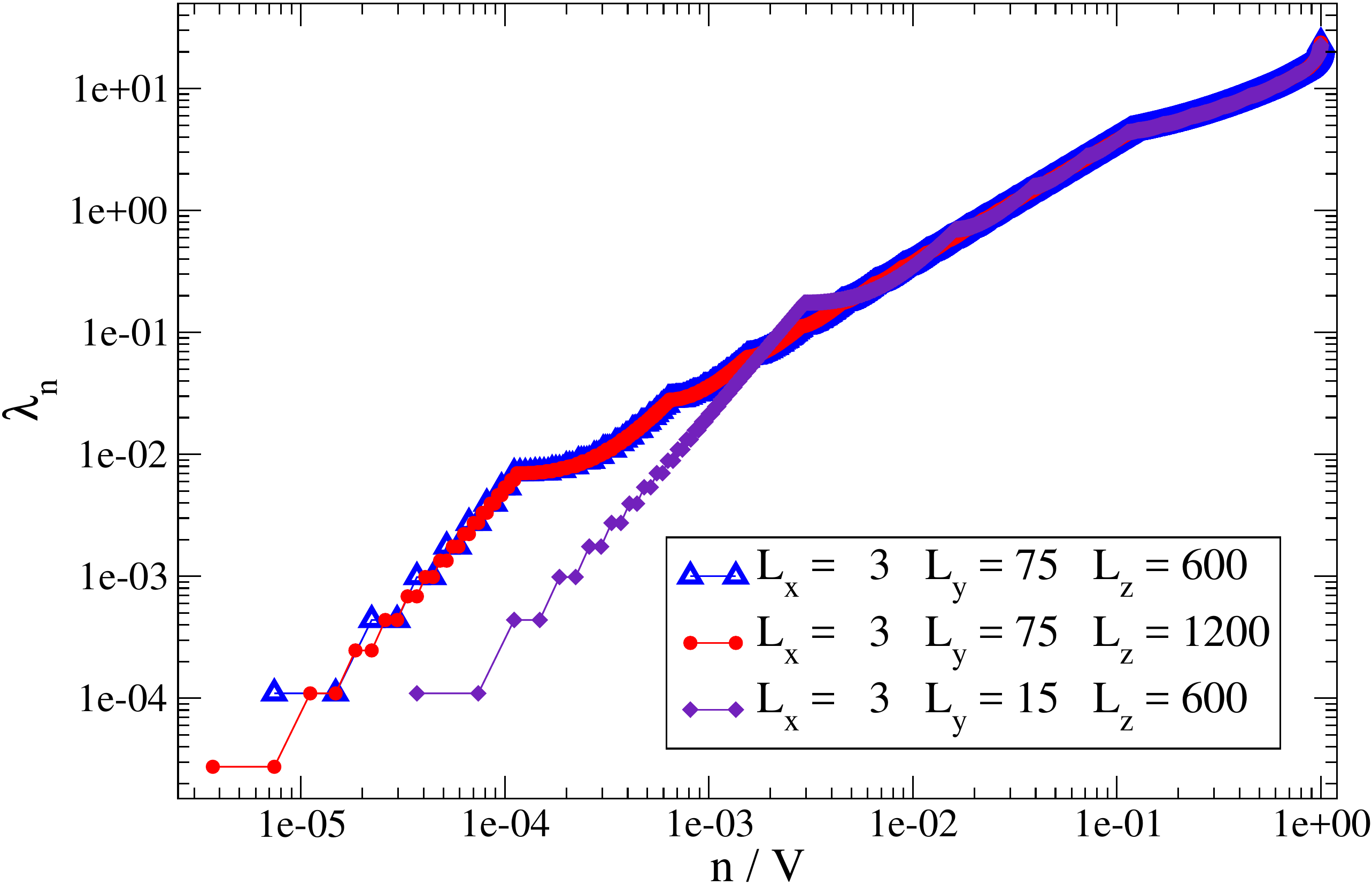}
 	\caption{Same as in Fig.~\ref{fig:toymodel_1}, for different 
combinations of the spatial sizes $L_i$ of the toroidal lattice. 
 	}
 	\label{fig:toymodel_2}
 \end{figure} 
 
 For the asymmetric lattices, instead, some well structured
 deviations emerge at low $n/V$, where $\lambda$ follows a 
 Weyl-like power law which is typical of lower dimensional models
 and can be easily interpreted as follows.
 For the lattice with $L_x = L_y = l = 15$ and $L_z = 600$, one
 does not find any eigenvalue with $m_x \neq 0$ and 
 $m_y \neq 0$ as long as 
 $\lambda < 4 \pi^2 / l^2 \simeq 0.175$, therefore in this range
 the distribution of eigenvalues is identical
 to that of a one-dimensional system, for which 
 $\lambda \propto (n/V)^{1/2}$; 
 for $\lambda > 4 \pi^2 / l^2$ also eigenvalues for which 
 $m_x$ and/or $m_y$ are non-zero appear, and their distribution 
 goes back to the standard 3-dimensional Weyl law.
 Making a wave-mechanics analogy, at low energy only longitudinal
 modes are excited, while transverse modes are frozen until
 a high enough energy threshold is reached. The point where
 one crosses from one power law behavior to the other
 brings information about the size of the shorter transverse scale.
 Similar considerations apply to the lattice 
 $L_x = 3$, $L_y = 75$ and $L_z = 600$, which has three
 different and well separated IR scales: in this case one 
 sees a one-dimensional power law for small $n/V$,
 which first turns into a two-dimensional one as modes
 in the $y$-direction start to be excited, and finally
 ends up in a standard 3d Weyl law when also modes
 with $m_x \neq 0$ come into play.

 The argument above can be rephrased at a more general level.
 Suppose we have a $D$-dimensional manifold where
 $d$ ``transverse'' dimensions are significantly shorter
 than the other $D -d$ ``longitudinal'' dimensions, 
 with a typical transverse scale 
 $l$. As long as one considers small eigenvalues,
 the modes in the transverse
 directions will not be excited, so that the counting of 
 eigenvalues will be given by 
 the Weyl law for $D - d$ dimensions, i.e.~$n(\lambda) = \omega_{D-d} (V/l^d) \lambda^{(D-d)/2} $.
 The change from one regime to the other will take place 
 when the transverse directions get excited for the first
 time, i.e.~at $\lambda \simeq \pi^2/l^2$ (the actual 
 prefactor depends on the details of the shorter 
 dimension), which corresponds to
 $n \propto V l^{-D}$, with a proportionality constant
 which depends only on the details of the short
 transverse scales and is independent of the 
 details of the longer scales. Therefore, different manifolds,
 sharing the same structure at short scales associated with 
 an effective dimensional reduction, lead to a distribution 
 $\lambda_n$ where the change from one power law behavior to the 
 other takes place at the same point in the 
 $(n/V)$-$\lambda$ plane, where $V$ is the global volume of the 
 manifold. The value of $n/V$, being proportional to $l^{-D}$, brings
 information about the size of the short scale.

 To better illustrate the concepts above, in Fig.~\ref{fig:toymodel_2}
 we show the distribution of 
 $\lambda_n$ as a function of $n/V$ for three different choices of 
 $L_x$, $L_y$ and $L_z$. The curves obtained for 
 $(L_x,L_y,L_z) = (3,75,600)$ and $(L_x,L_y,L_z) = (3,75,1200)$
 go exactly onto each other: their short scale structure is the same
 and the function $n(\lambda)$ just differs for different number
 of modes which are counted along the large direction $L_z$, however
 this difference disappears when one considers the scaling variable
 $n/V$, leading to a perfect collapse. The collapse instead is not perfect
 when one considers the lattice 
 $(L_x,L_y,L_z) = (3,15,600)$, which has a different ``intermediate'' scale:
 moving from large to small $n/V$, the turning point from dimension 
3 to dimension 2 
 is the same as for the two other lattices, however the turning point
 from dimension 2 to dimension 1 takes place earlier, because $L_y$ is shorter.
 \\
 
 The possible examples which one can discuss within the toy model
 are quite limited. For instance, one cannot consider the case in which 
 there are points where the manifold branches into multiple connected 
 ramifications, something which in general can lead to an increase,
 instead of a decrease, of the effective dimension. However, extrapolating
 the arguments given above, we can conjecture the following.
 $D$-dimensional manifolds having different overall volumes and shape, 
 but sharing a similarity in the structures which are found at intermediate and 
 short scales, will lead to similar (i.e.~collapsing onto each other)
 curves when $\lambda_n$ is plotted against $n/V$, $V$ being the total
 volume of the manifold. Moreover, the power law taking place
 at a given value of $n/V$ will give information about the effective 
 dimensionality $d_{EFF}$ of the manifold at a scale
 of the order $(n/V)^{-1/D}$, with
 \begin{equation}
 \frac{2}{d_{EFF}} = \frac{d \log \lambda}{d \log (n/V)} \, .
 \label{eq:weyl_dimension} 
 \end{equation}
 This kind of information is similar to what is obtained by implementing
 diffusive processes to measure the spectral
 dimension.
 
  \section{Numerical Results}
  \label{sec:numres_eigenvalues}
  
  In this section we present results 
  regarding mostly the spectrum of the LB operator defined 
  on spatial slices, while a detailed discussion 
  regarding the eigenvectors is postponed 
  to a forthcoming study.
  We performed the analysis on spatial slices of configurations in each phase; 
  in particular, almost all the results shown come from simulations running 
  deep into each phase, at the points circled and labeled by a letter in 
  Fig.~\ref{fig:phasediag} and in Table~\ref{tab:simpoints}. 

  \begin{figure}[t!]
  	\centering
  	\includegraphics[width=1\columnwidth]{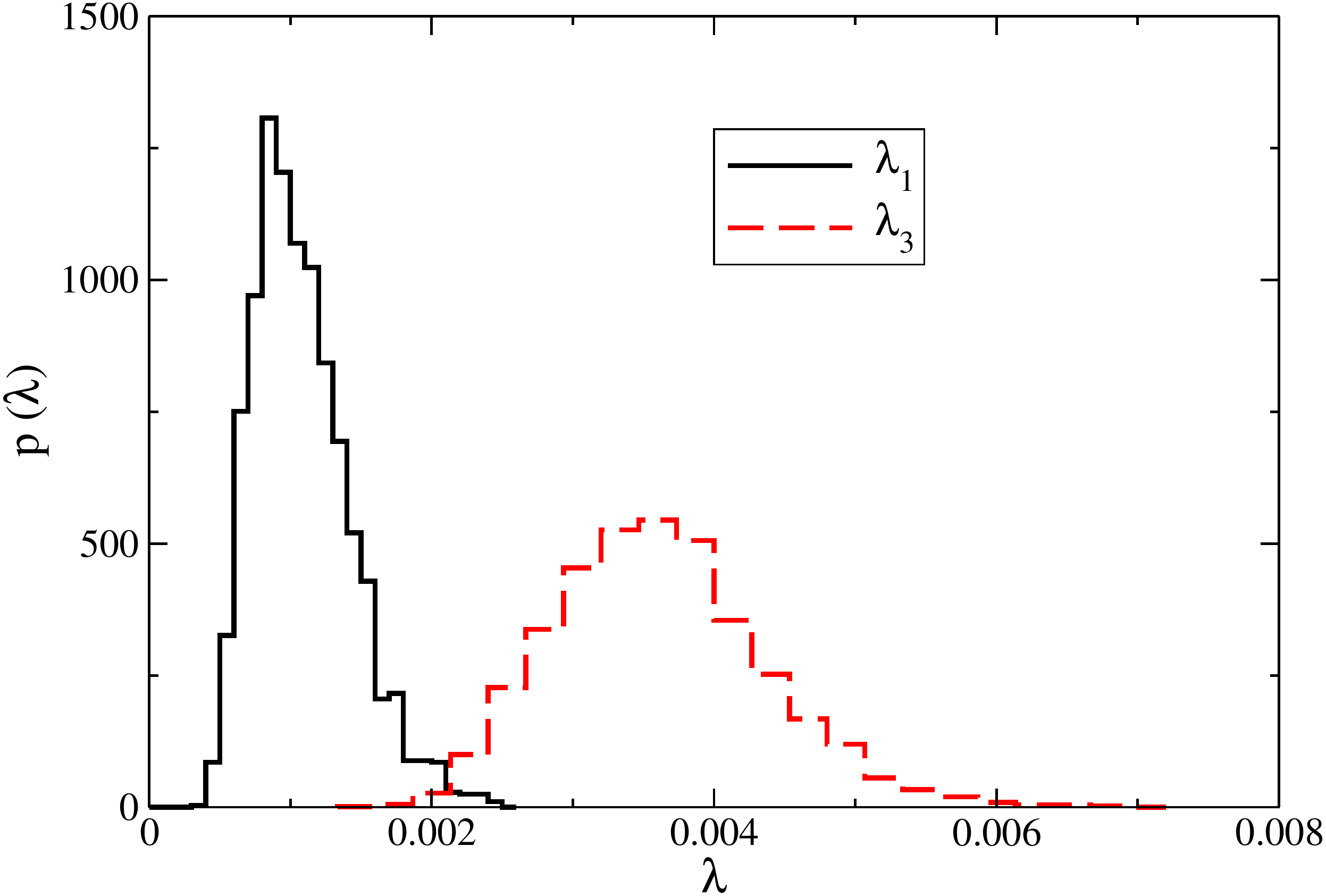}
  	\caption{Probability distribution of $\lambda_1$ and $\lambda_3$
  		for slices with $V_S \simeq 2300$, taken from configurations sampled deep in
  		the $C_{dS}$ phase (simulation point $c$), and with total spatial volume $V_{S,tot}=\frac{N_{41}}{2} = 40 k$.
  	}
  	\label{fig:dis_l1_l3}
  \end{figure}

  While the total spatial volume has been fixed in 
  each simulation to a target value, the spatial 
  volume of single slices, $V_S$, can vary greatly from 
  one slice to the other (apart from phase $B$).
  That will permit us to access the dependence of the 
  spectrum on $V_S$, an information that will
  be very important for many aspects.
  As discussed above, each spatial slice will be associated with
  a 4-regular undirected graph, with each vertex of the graph 
  corresponding to a spatial tetrahedron.
  For this reason, it will be frequent in the following discussion
  to borrow concepts and terminology from graph theory.
  
  We will first look at the low lying part of the spectrum,
  show how the transition from one phase to the other can be associated 
  to the emergence of a gap in the spectrum, and discuss what that means
  in terms of the geometrical properties of the triangulations.
  We will then turn to results regarding the whole spectrum
  and show how one can obtain information on the effective 
  dimension of the geometry at different scales. 
  Finally, we will describe two methods to visualize graphs and apply
  them to show the appearance of spatial slices.

  \begin{figure}[t!]
  	\centering
  	\includegraphics[width=1\columnwidth]{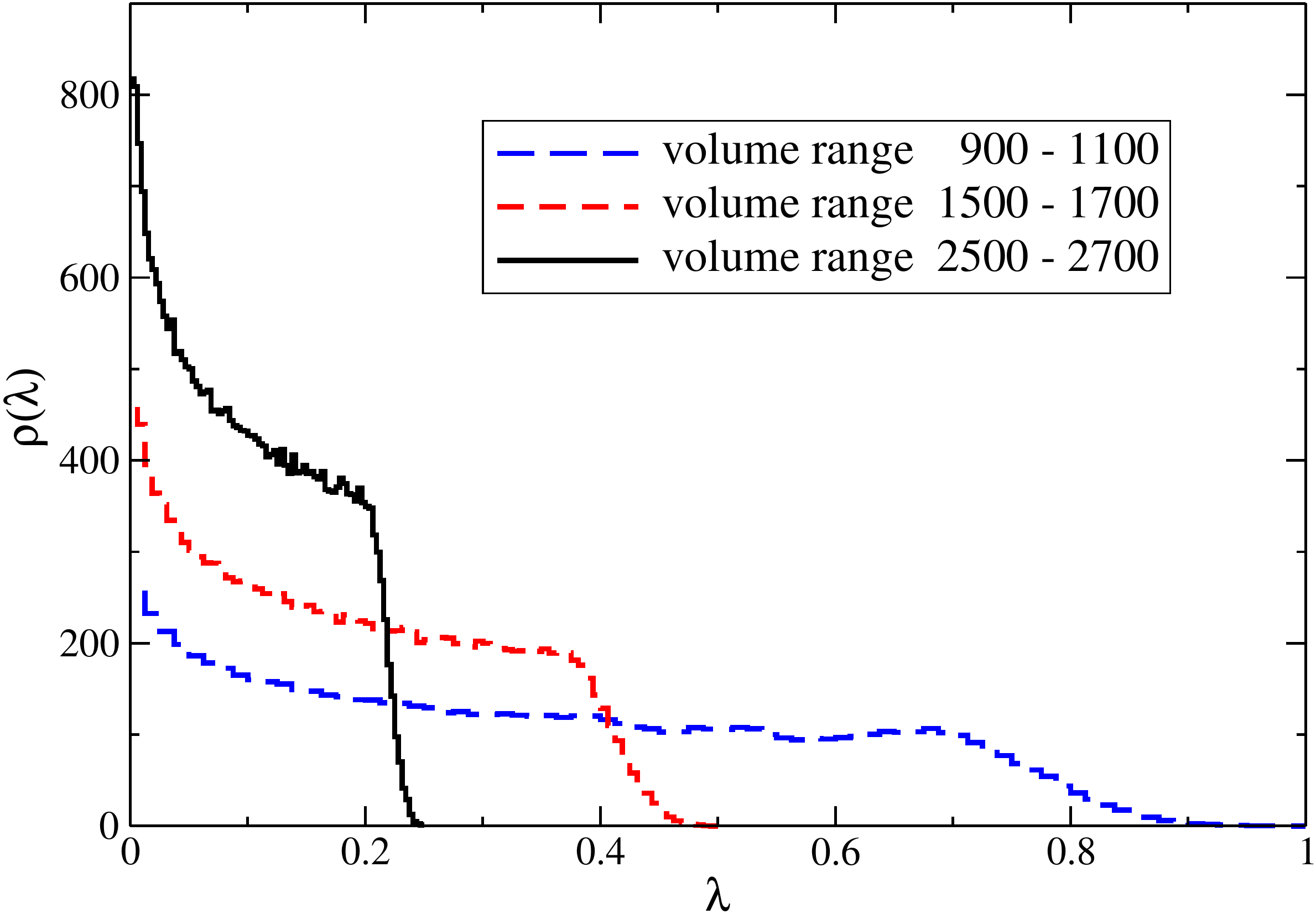}
  	\caption{Density $\rho(\lambda)$ 
computed from the first 100 eigenvalues for slices deep in the $C_{dS}$ phase (simulation point $c$) with total spatial volume $V_{S,tot} = 40k$, and for different ranges of the spatial slice volume $V_S$. 
 	}
  	\label{fig:CdS_100eigs}
  \end{figure}
  
  \subsection{The low lying spectrum and the emergence of a gap}
  \label{subsec:low_spect}

  Apart from the zero eigenvalue, $\lambda_0 = 0$, 
  the remaining eigenvalues will fluctuate randomly 
  from one configuration to the other and, moreover, their distribution
  will depend on $V_S$ in a well defined way 
  that we are going to discuss later on.
  As an example, in Fig.~\ref{fig:dis_l1_l3} we show the distribution 
  of $\lambda_1$ and $\lambda_3$ on a set of around $3 \times 10^3$ slices of approximately equal volume $V_S \simeq 2300$ and in $C_{dS}$ phase.
  Therefore, while the spectrum of each spatial slice 
  is intrinsically discrete, because of the finite number of 
  vertices making up the associated graph, it makes sense
  to define a continuous distribution $\rho(\lambda)$,
  assigned so that $\rho (\lambda) d \lambda$ gives back the number
  of eigenvalues which are found on average in the interval 
  $[\lambda, \lambda + d \lambda]$. 
  In general
  $\rho(\lambda)$ will be a function of the bare 
  parameters chosen to sample the triangulations and,
  for fixed parameters, of the spatial volume $V_S$ 
  of the chosen slice.

  \begin{figure}[t!]
  	\centering
  	\includegraphics[width=1\linewidth]{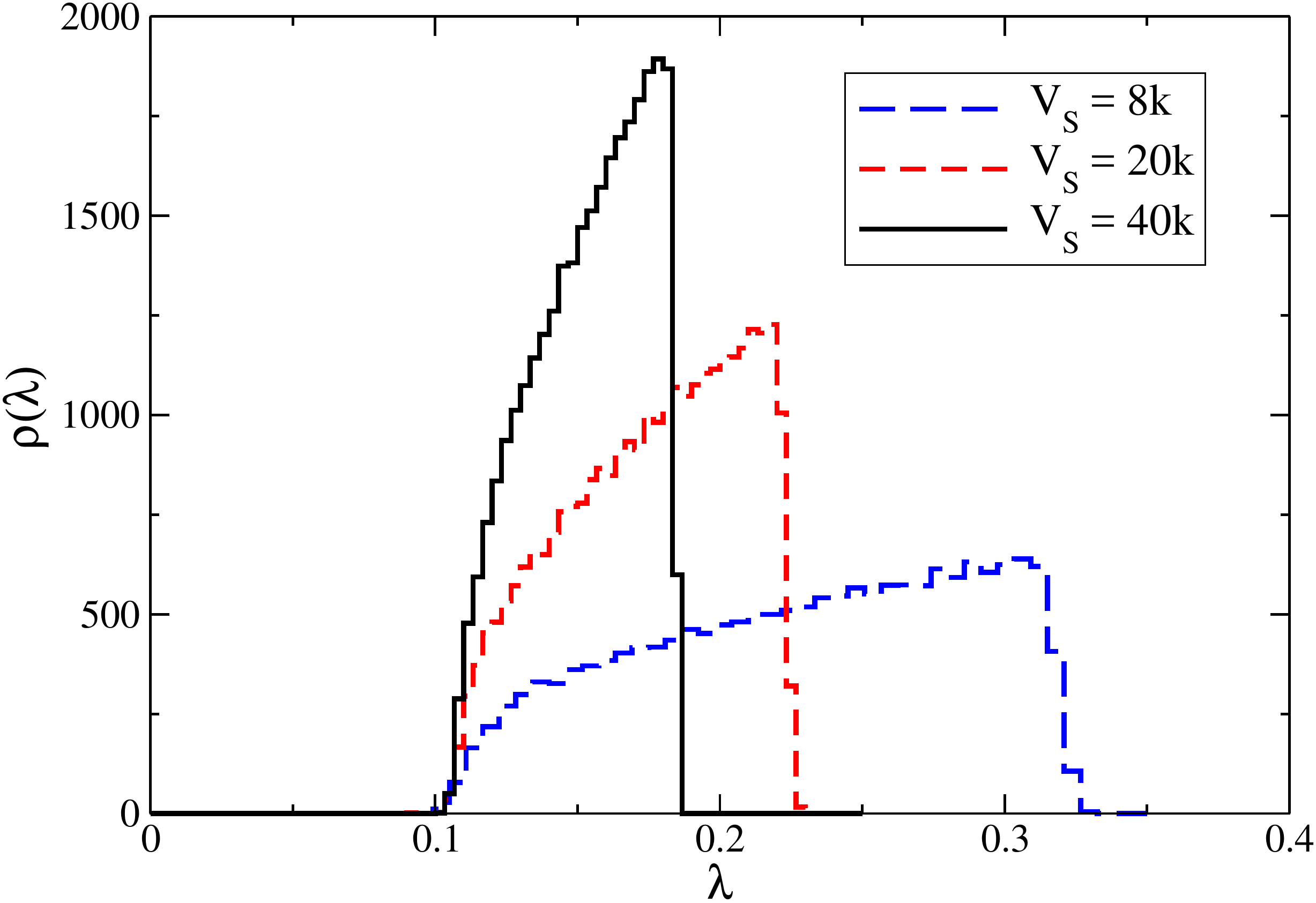}
  	\caption{
Density $\rho(\lambda)$ computed from the first 100 eigenvalues
for the maximal slices in the $B$ phase (simulation point $b$) and for different spatial volumes $V_S$. 
}
  	\label{fig:B_100eigs}
  \end{figure}

  In Figs.~\ref{fig:CdS_100eigs} and \ref{fig:B_100eigs} 
  we show the low lying part 
  of the distribution $\rho(\lambda)$ obtained from simulations
  performed respectively in the $C_{dS}$ and $B$ phases, selecting
  in each case three different ranges of spatial volumes\footnote{For the 
  	$B$ phase, different spatial
  	volumes correspond actually to different simulations
  	with different constraint on $N_{41}$, since 
  	in this case most of the spatial volume is contained in one single slice}. 
  In order to focus just on the low part of the spectrum,
  we have limited the input for $\rho$ to just the first few eigenvalues
  in each case ($n \leq 100$).
  
  A striking difference between the two phases emerges. 
  In the $B$ phase there is a gap
  $\Delta \lambda = \lambda_1 \simeq 0.1$ 
  which does not disappear and is practically constant 
  as the spatial volume $V_S$ grows, i.e.~as one approaches the thermodynamical
  limit. This gap is absent in the $C_{dS}$ phase, where the distribution
  of the first 100 eigenvalues is instead more and more squeezed towards
  $\lambda = 0$ as $V_S$ grows. The presence or absence of a gap
  in the spectrum is a characteristic which distinguishes different
  phases in many different fields of physics: think for instance of 
  Quantum Chromodynamics, where the absence/presence of a gap in the 
  spectrum of the Dirac operator distinguishes between the phases with 
  spontaneously broken/unbroken chiral symmetry. Let us discuss what is 
  the meaning of the gap in our context.

  Graphs which maintain a finite gap as the number of vertices goes to infinity 
  are known as {\em expander graphs}~\cite{expgraphs} and play a significant role
  in many fields, e.g., in computer science. They
  are characterized by a high connectivity, i.e.~the boundary of every 
  subset of vertices is generically large. Such a high connectivity
  is usually associated with a degree of randomness, i.e.~lack of order,
  in the connections between vertices: for instance, random regular graphs 
  are {\em expanders} with high probability~\cite{alon_second_conjecture}. 
The strict relation between the high connectivity and the presence of 
a finite gap in the spectrum is also encoded in 
Cheeger's inequalities, see Eq.~(\ref{eq:Cheeger_ineq}).

The property which is maybe most relevant
  to our context is the fact that the diameter of an expander,
  defined as the maximum distance\footnote{The distance 
  	between a pair of vertices is defined as the length
  	of the shortest path (i.e.~the geodesic) connecting the two vertices.}
  between any pair of vertices,
  does not grow larger than logarithmically with total number of 
  vertices~\cite{diameters_and_eigenvalues,diameter_lower_bound}. 
  Therefore, in this phase the spatial slices 
  do not develop a well defined geometry, since 
  the size (diameter) of the Universe
  remains small as the volume tends to infinity,
a fact described also in previous CDT studies in terms
of a diverging Hausdorff dimension.
  This fact 
  can be easily interpreted in terms of diffusive processes: 
  as argued above (see Section~\ref{sec:LBproperties}), 
  the value of the spectral gap, $\lambda_1$, can be interpreted as the inverse 
  of the diffusion time of the slowest mode; the fact that
  the time to diffuse through the whole Universe stays finite
  means that its size is not growing significantly.
  
  \begin{figure}
  	\centering
  	\includegraphics[width=1.0\linewidth]{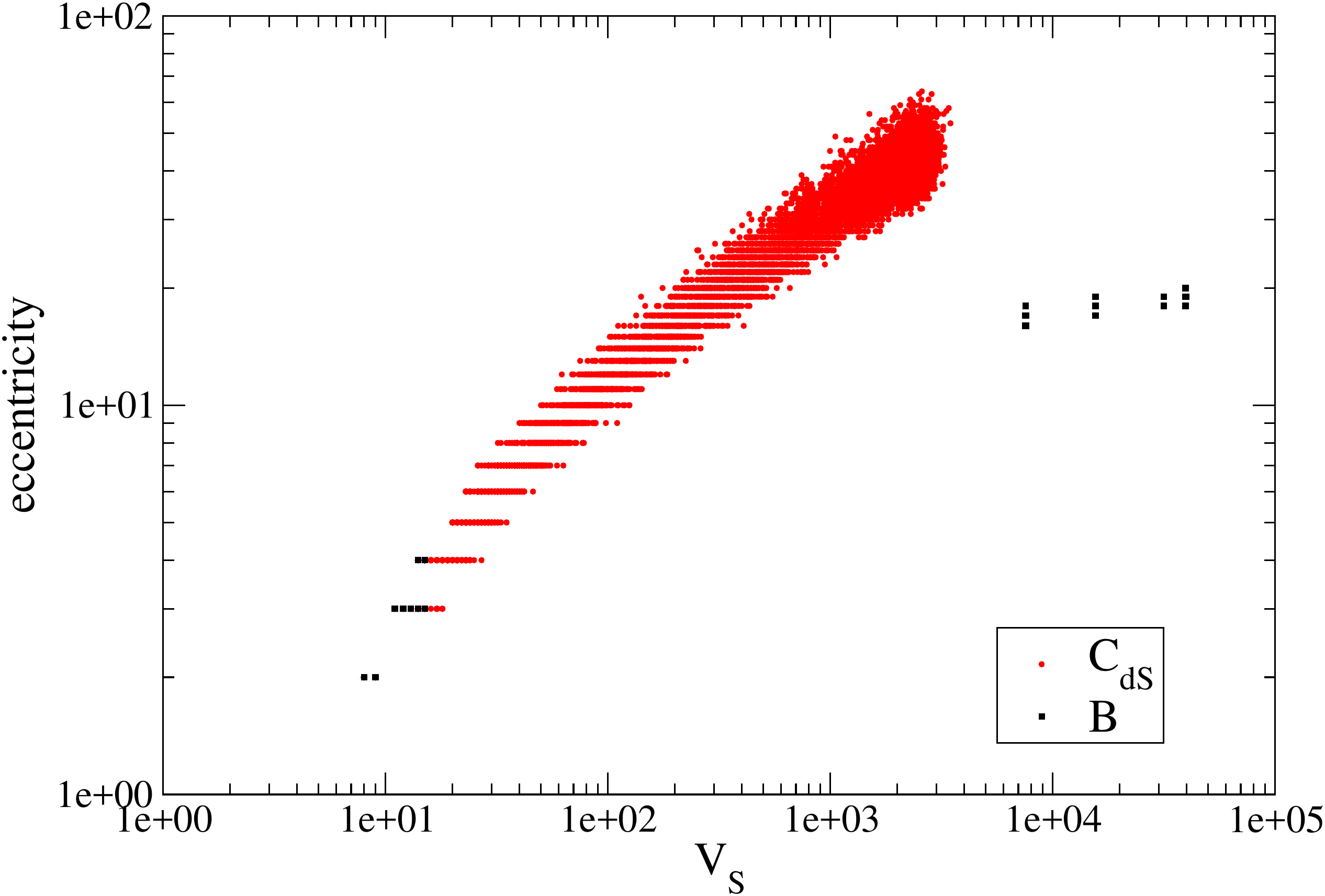}
  	\caption{Scatter plot of the eccentricity of $200$ randomly selected vertices
  		for each slice of about $400$ configurations in the $C_{dS}$ phase (simulation point $c$) with total spatial volume $V_{S,tot}=20k$, and for the maximal slices of about $200$ configurations in the $B$ phase (simulation point $b$) with total volumes $V_{S,tot}=8k,16k,32k,40k$. Results are reported against the slice volume $V_S$.}
  	\label{fig:diam_CdS_B_ecc-vs-V}
  \end{figure}

  On the contrary, according to the arguments
  discussed in Section~\ref{subsec:toymodel},
  for a graph representing a 
  standard manifold having a finite
  effective dimension on large scales, one expects 
  that the number of eigenvalues found 
  below any given $\lambda$ should grow proportionally
  to the volume $V_S$, 
  $n \propto V_S\, \lambda^{d_{EFF/2}}$, 
  see for instance  
  Eq.~\eqref{eq:weyl_dimension}. That means that the 
  gap must go to zero as $V_S \to \infty$
  and, moreover, that a finite normalized density\footnote{A finite 
  	density of eigenvalues around $\lambda = 0$ is a 
  	condition stronger than the simple absence of a gap. Indeed,
  	one might have situations in which isolated quasi-zero
  	eigenvalues develop, while the continuous part of the 
  	spectrum maintains a gap: think for instance of two expander
  	graphs connected by a thin bottleneck.} of eigenvalues,
  $\rho(\lambda)/V_S$, must develop around $\lambda = 0$.
  Instead, as it will be shown in more detail below, 
  the presence of a spectral gap for 
  slices in the $B$ phase indicates that the effective dimension is indeed 
  diverging at large scales, in agreement with the high connectivity 
  property.
  
  As an independent check, we computed the maximum distance from a randomly chosen vertex to 
  all other vertices in the graph (a quantity usually called the \emph{eccentricity} of the vertex), 
  iterating the procedure for $200$ different starting vertices and for each slice in the $C_{dS}$ 
  and $B$ phases. The maximum eccentricity in a graph corresponds to its diameter, 
  so the eccentricity of a random vertex is actually a lower bound to the diameter.
  Therefore the results, which are shown in the form of a scatter plot in 
  Fig.~\ref{fig:diam_CdS_B_ecc-vs-V}, are consistent with a diameter which, for sufficiently large volumes, grows as a
  power law of $V_S$ in phase $C_{dS}$, while on the contrary it seems to reach a constant
  or to grow at most logarithmically in the $B$ phase.

\begin{figure}
  	\centering
  	\includegraphics[width=1\linewidth]{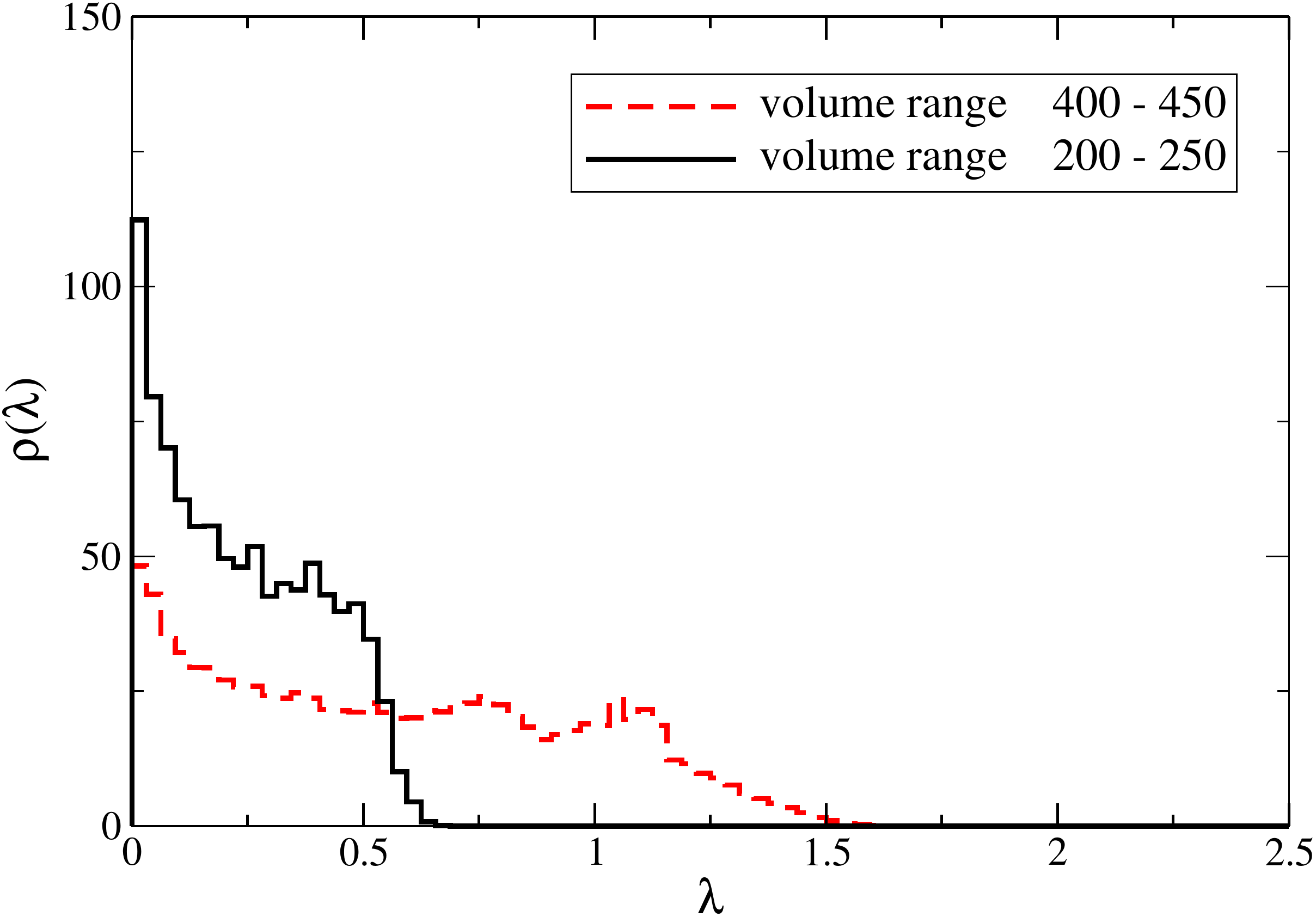}
  	\caption{
  		Density $\rho(\lambda)$ computed from the first 30 eigenvalues
  		for slices deep in the $A$ phase (simulation point $a$) with total spatial volume $V_{S,tot} = 8k$, and for two different ranges of the spatial slice volume $V_S$. 
}
  	\label{fig:A_100eigs}
  \end{figure}

  The properties of slices in phase $A$ are quite similar to those 
  found in phase $C_{dS}$, i.e.~one has evidence for a finite 
  density of eigenvalues around $\lambda = 0$ in the large
  $V_S$ limit, even if the distribution of slice volumes is 
  significantly different from that found in phase $C_{dS}$. 
  An example of the distribution of the first 
  30 eigenvalues in this phase is reported in Fig.~\ref{fig:A_100eigs}.
  
\begin{figure}
  	\centering
  	\includegraphics[width=1\linewidth]{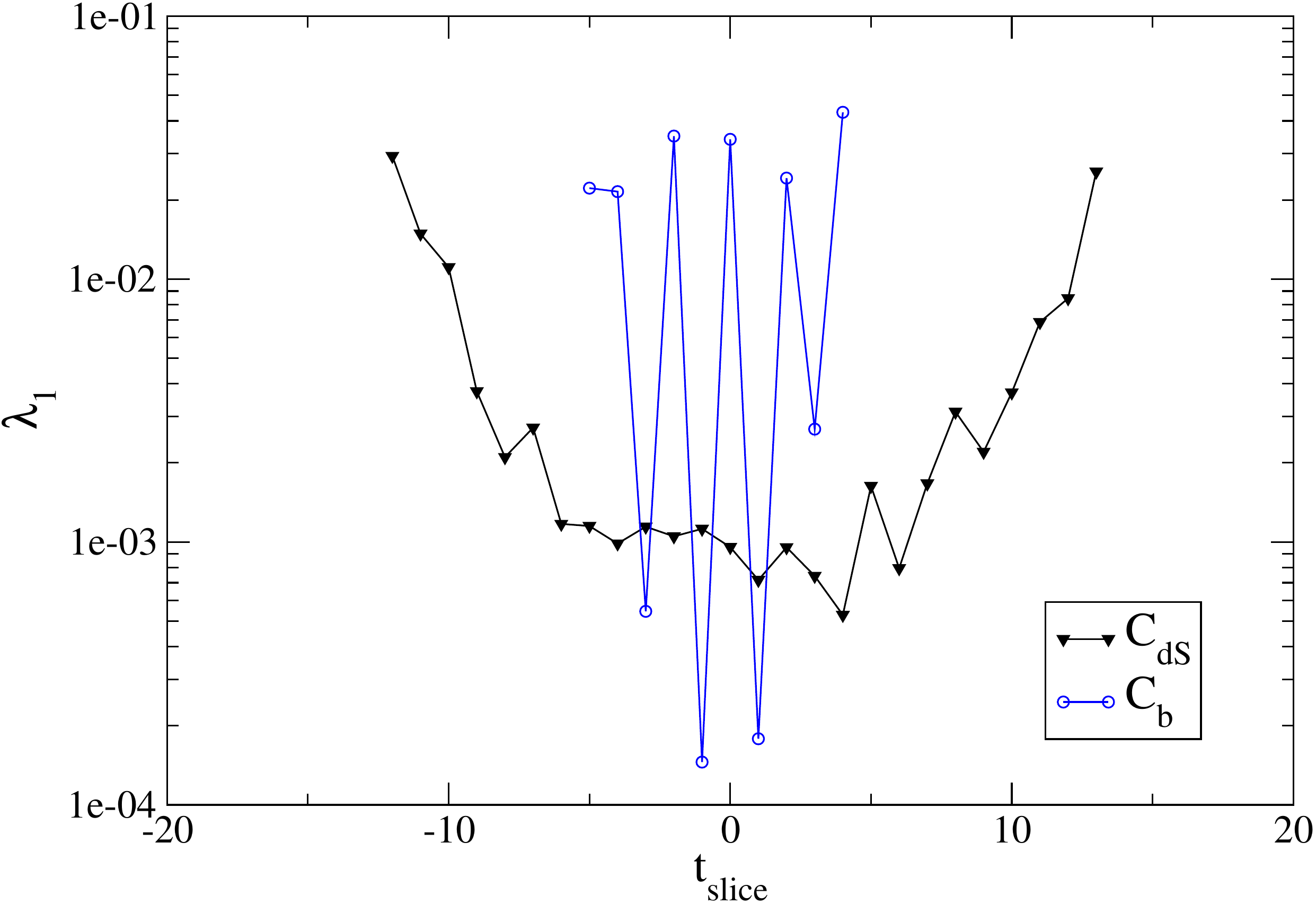}
  	\caption{Spectral gap $\lambda_1$ as a function of the slice-time for single configurations in $C_b$ and $C_{dS}$ phases with total spatial volume $V_{S,tot}=40k$ and with the slice-time of maximal slice shifted to zero. Only slices in the bulk (with volume $V_S\ge 200$) have been shown.}
  	\label{fig:single_Cb-CdS_lam1-tslice}
  \end{figure}

\begin{figure}
  	\centering
  	\includegraphics[width=1\linewidth]{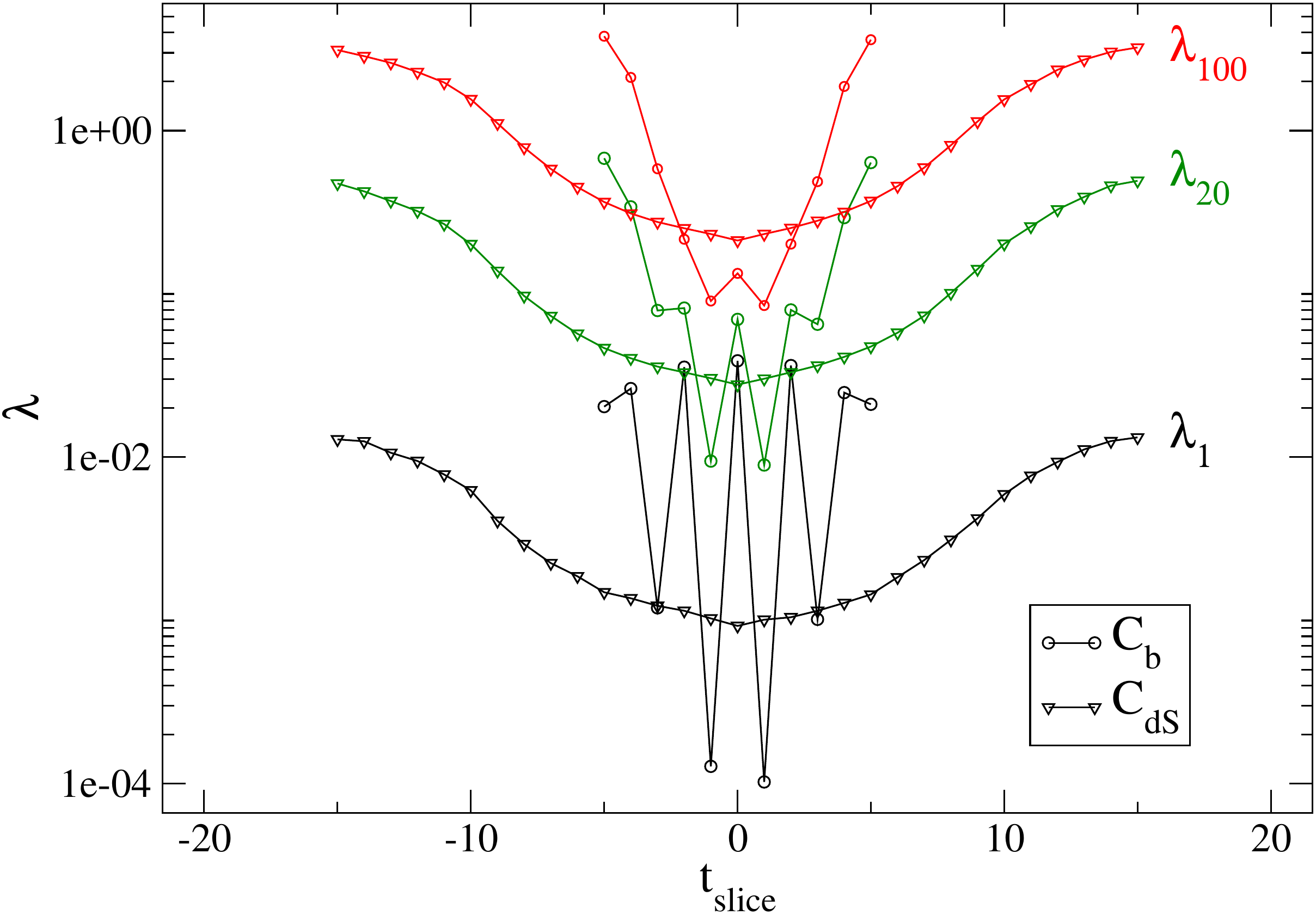}
  	\caption{Averages of $\lambda_1$,$\lambda_{20}$ and $\lambda_{100}$ as a  function of the slice-time for configurations in $C_b$ and $C_{dS}$ phases, where the slice-time of maximal slices has been shifted to zero. Only slices in the bulk (with volume $V_S\ge 200$) have been shown.}
  	\label{fig:aver_Cb-CdS_lam1-20-100_vs_tslice}
  \end{figure}

  Instead, the spectra of slices in the bifurcation phase $C_b$ 
  need a separate treatment.
  Indeed, it is well known that 
  the bulk of the configurations are made up of two 
  separate classes of slices, which alternate 
  each other in slice-time
  and have different properties~\cite{cdt_newhightrans}: it is 
  reasonable to expect that this is reflected
  also in their spectra.
  This is indeed the case, as can be appreciated 
  by looking at Fig.~\ref{fig:single_Cb-CdS_lam1-tslice},
  where we report the value of $\lambda_1$
  obtained on the different slices (i.e.~at different Euclidean times)
  for a typical configuration sampled in the
  $C_b$ phase, and compare it to a similar plot obtained
  for the $C_{dS}$ phase. For an easier comparison, 
  the time coordinates of the slices have been relabeled 
  in each case so that the slice with the 
  largest volume corresponds to $t_{slice} = 0$; moreover, 
  we restricted to the bulk of configurations (i.e.~we chose slices with $V_S>200$).
  Contrary to the $C_{dS}$ phase, in the $C_b$ phase $\lambda_1$ changes abruptly
  from one slice to the other, with small values 
  alternated with larger ones, differing by even two order
  of magnitudes. 
  This striking difference, which emerges even for single configurations,
  is even more clear as one considers the whole ensemble:
  Fig.~\ref{fig:aver_Cb-CdS_lam1-20-100_vs_tslice} shows the average of $\lambda_1$, $\lambda_{20}$ and $\lambda_{100}$ for configurations in the $C_{b}$ and $C_{dS}$ phases, with slice times relabeled as before.
  In the $C_{dS}$ phase $\lambda_1$ changes smoothly 
  with $t_{slice}$, and this change is mostly induced
  by the corresponding change of the slice volume,
  while in the $C_b$ phase 
  the alternating structure is visible also
  for higher eigenvalues, even if somewhat reduced and limited
  to the central region as $n$ grows.
  
  Therefore, we conclude that 
  the alternating structure of spatial slices is apparent and 
  well represented in the low-lying spectra: slices in the bulk of $C_b$ phase configurations can 
  be separated in two distinct classes by the value of their spectral gap, while in the $C_{dS}$ 
  phase there is no sharp distinction apart from a volume-dependent behavior connected to an 
  observed Weyl-like scaling, which will be discussed in more detail 
  in Section~\ref{subsec:scalings}.
  
\begin{figure}
  	\centering
  	\includegraphics[width=1\linewidth]{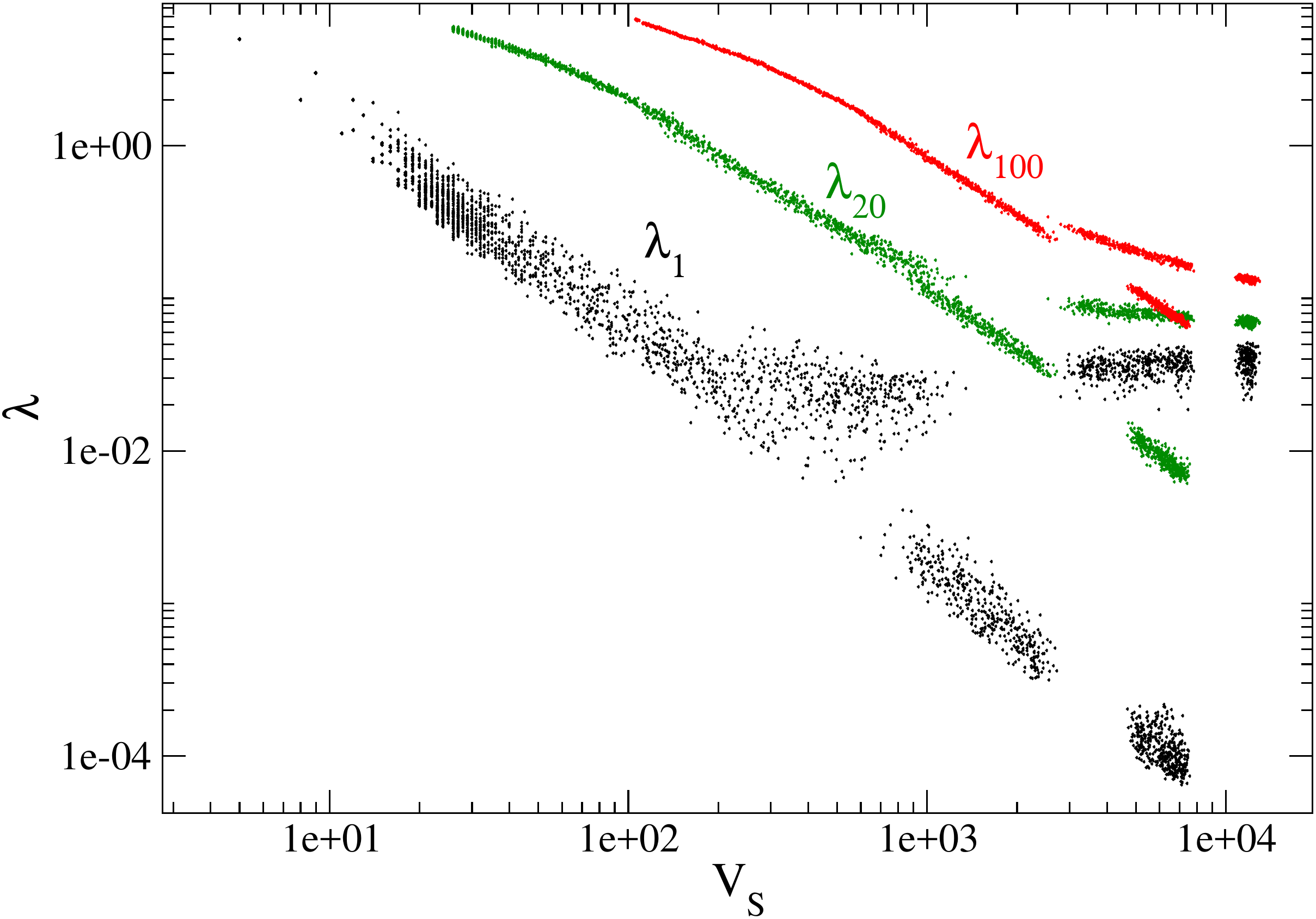}
  	\caption{Scatter plot of the values of $\lambda_1$, $\lambda_{20}$ and $\lambda_{100}$ versus the volume of the slice on which they are computed, for slices of configurations deep in the $C_b$ phase (simulation point $\widetilde{c}$) and with volume fixing $V_S = 40k$.}
  	\label{fig:dt4_lam1-20-100_vs_V}
  \end{figure}

  In order get a better perspective on these results, in Fig.~\ref{fig:dt4_lam1-20-100_vs_V} we show the eigenvalues $\lambda_n$, 
  with $n=1,20,100$, plotted against the volume of the slice on which they are computed, for the slices of all configurations 
  sampled in the $C_b$ phase (in particular 
  at the simulation point labeled $\widetilde{c}$). 
  Slices with volumes larger than a given $V_S$, which we call
  \emph{bifurcation volume}\footnote{It is interesting no notice that
the $C_b$ phase can be called {\em bifurcation phase} for many different reasons.}, divide in two distinct classes 
  characterized by $\lambda_n$ taking values in well separated ranges.
  It is interesting that such bifurcation volume depends on $n$: 
  that also explains why in Fig.~\ref{fig:aver_Cb-CdS_lam1-20-100_vs_tslice} the alternating behavior of higher order eigenvalues (e.g., $\lambda_{100}$) drops off earlier than lower order ones, since spatial volumes get smaller far from the slice with maximal volume and then their volume becomes less than the bifurcation one at that order. That actually means that the alternating slices
found in the $C_b$ phase only differ for the low lying part of the LB spectrum, 
while for high enough eigenvalues order they are not distinguishable; high eigenvalues 
mean small scales, hence we expect that the alternating slices have the same
small scale structures and only differ at large scales. We will come back on this point
later on.

  In view of the close similarity with the properties of slices found respectively 
  in the $C_{dS}$ in the $B$ phase, we assign to the two classes of slices the
  names $dS$-type (low spectral gap) and $B$-type (high spectral gap). 
  Looking again at Fig.~\ref{fig:dt4_lam1-20-100_vs_V}, we notice 
  that, for sufficiently large volumes, the two classes populate only specific volume ranges.
  Furthermore, the maximal slice in the $C_b$ phase is typically observed to be of $B$-type, with a volume 
  ranging in a narrow interval which is separated from 
  the volumes of the other slices. This alternating distribution
of spatial volumes has been indeed one of the first signals of the presence of 
the new phase~\cite{Ambjorn:2014mra,Ambjorn:2015qja}.
  
\begin{table}[b!]
{\large
 	\centering
 	\begin{tabular}{ | >{\centering}m{2cm} | >{\centering}m{2cm} | >{\centering}m{2cm} | >{\centering}m{2cm} |}
\hline 		$A$	& $B$	&  $C_b$ & $C_{dS}$ \tabularnewline \hline
 		no-gap	&	gap	&	gap ~ no-gap	&	no-gap \tabularnewline
\hline 
	\end{tabular}
} 	
\caption{Characterization of the phase diagram of CDT according to the 
   zero or non-zero gap of the LB operator as an order parameter.}
\label{tab:gap}
 \end{table}

\begin{figure}
  	\centering
  	\includegraphics[width=1.1\linewidth]{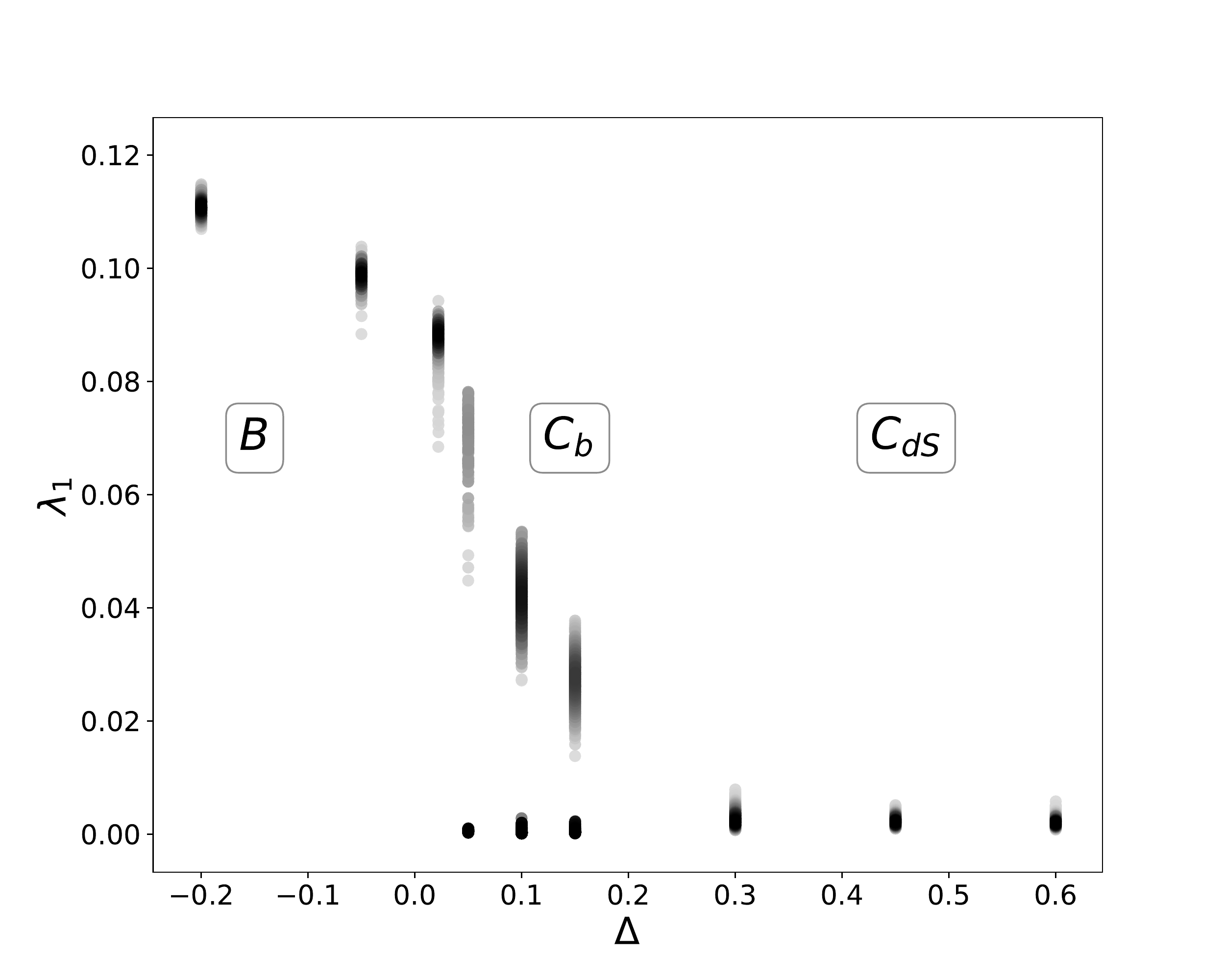}
  	\caption{Distribution of $\lambda_1$ (in scatter plot format) for $k_0=2.2$ and variable $\Delta$ for configurations with total spatial volume $V_{S,tot}=\frac{N_{41}}{2}=40k$ and considering only slices with spatial volume $V_S > 2k$.}
  	\label{fig:lambda1_op-scatt}
  \end{figure}

  A summary of the characterization of the phase diagram of CDT according to the 
   (zero or non-zero) gap of the LB operator as an order parameter 
   is reported in Table~\ref{tab:gap}.
To conclude the discussion about the gap, it is interesting to consider how the distribution of 
  $\lambda_1$ changes across the different phases. 
  To this purpose, in Fig.~\ref{fig:lambda1_op-scatt} we show a scatter
  plot of $\lambda_1$ for different values of $\Delta$ at fixed
  $k_0=2.2$: darker points corresponds to more frequent values of $\lambda_1$.
  As $\Delta$ increases, the gap in the $B$ or $B$-type slices progressively reduces and 
  approaches zero at the point where one enters the $C_{dS}$ phase. A
  	gap in the spectrum is a quantity which has mass dimension two (as the LB operator), i.e.~an inverse length squared:
  	if future studies will show that the drop to zero takes place in a continuous way,
  	that will give evidence for a second order phase transition with a diverging correlation length.

  \subsection{Scaling and spectral dimension}\label{subsec:scalings}
  
\begin{figure}
  	\centering
  	\includegraphics[width=1\linewidth]{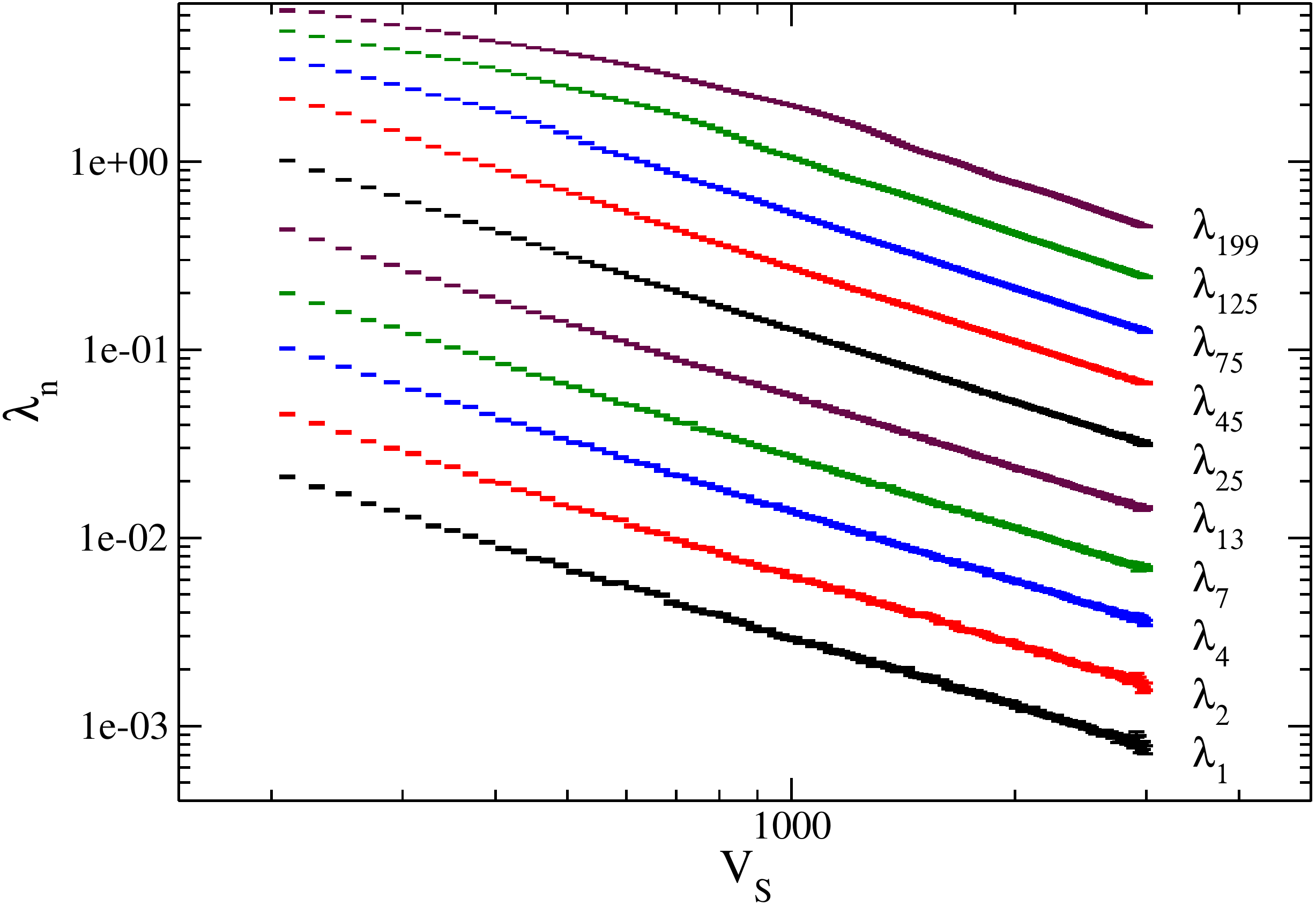}
  	\caption{Averages of eigenvalues $\lambda_n$ for selected orders $n$ and computed in narrow bins of volumes ($\Delta V_S = 20$), for slices of configurations sampled deep into the $C_{dS}$ phase (simulation point $c$), with total spatial volume $V_{S,tot}=40k$.}
  	\label{fig:avereig_vs_vols}
  \end{figure}

  As one expects, and as it emerges from some of the results that we have 
  already shown, the typical values
  obtained for the $n$-th eigenvalue of the LB operator on spatial
  slices, $\lambda_n$, scale with 
  the volume $V_S$ of the slice, and in a different way for the different phases.
  As an example, 
  in Fig.~\ref{fig:avereig_vs_vols} we show the average values obtained for $\lambda_n$ 
  (for a few selected values of $n$), as a function of the volume, in the $C_{dS}$ phase.

  \begin{figure}
  	\centering
  	\includegraphics[width=1\linewidth]{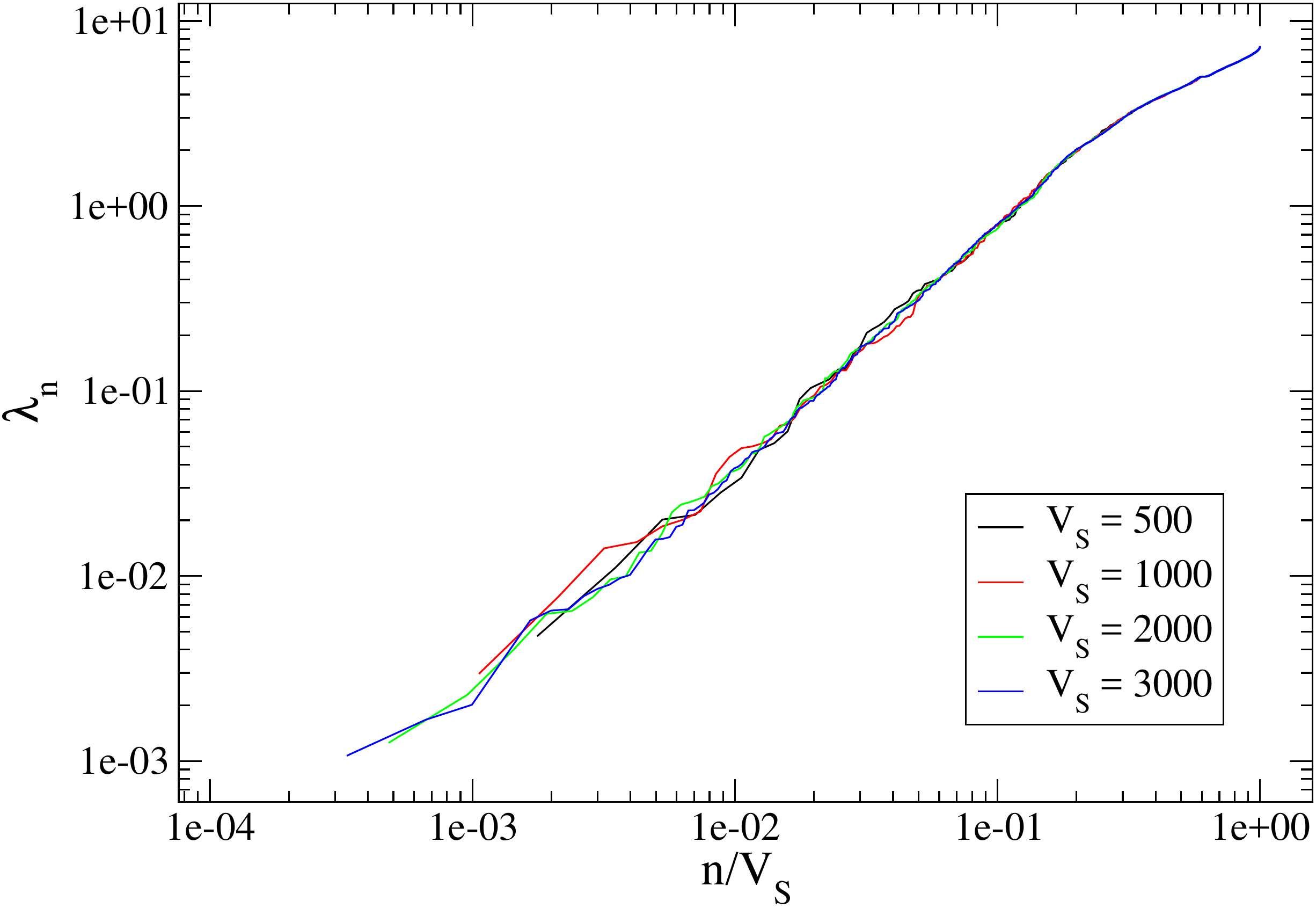}
  	\caption{Plot of $\lambda_n$ against its 
volume-normalized order $n/V_S$, for four randomly selected slices with volumes $V_S \simeq 500,1000,2000,3000$, taken 
from  configurations sampled deep into the $C_{dS}$ phase (simulation point $c$) with total spatial volume $V_{S,tot}=40k$.}
  	\label{fig:plot_CdS_lam-k_V_collapse}
  \end{figure}
  
  In order to better interpret this scaling, and inspired by the discussion 
  reported in Section~\ref{subsec:toymodel}, in the following we will consider
  how $\lambda_n$ depends on the variable $n/V_S$. To show that this may indeed be illuminating,
  in Fig.~\ref{fig:plot_CdS_lam-k_V_collapse} we report $\lambda_n$ as a function of $n/V_S$ 
  for four spatial slices, which have been 
  randomly picked from an ensemble produced in the 
  $C_{dS}$ phase and have quite different
  volumes, ranging over almost one order of magnitude\footnote{Notice that $n/V_S$ can take values in the range $(0,1)$ 
  	(recall that $\lambda_0=0$ is excluded from our discussion), while the maximum eigenvalue 
  	$\lambda$ is always bounded by $2k = 8$, that is twice the degree of vertices in the $k$-regular 
  	graph.}. 
  The collapse of the four curves onto each other is impressive and, in view of the discussion
  in Section~\ref{subsec:toymodel}, can be interpreted in this way: despite the fact 
  that the slices have quite different extensions, they show the same
  kind of structures at intermediate common scales.
  
\begin{figure}
  	\centering
  	\includegraphics[width=1\linewidth]{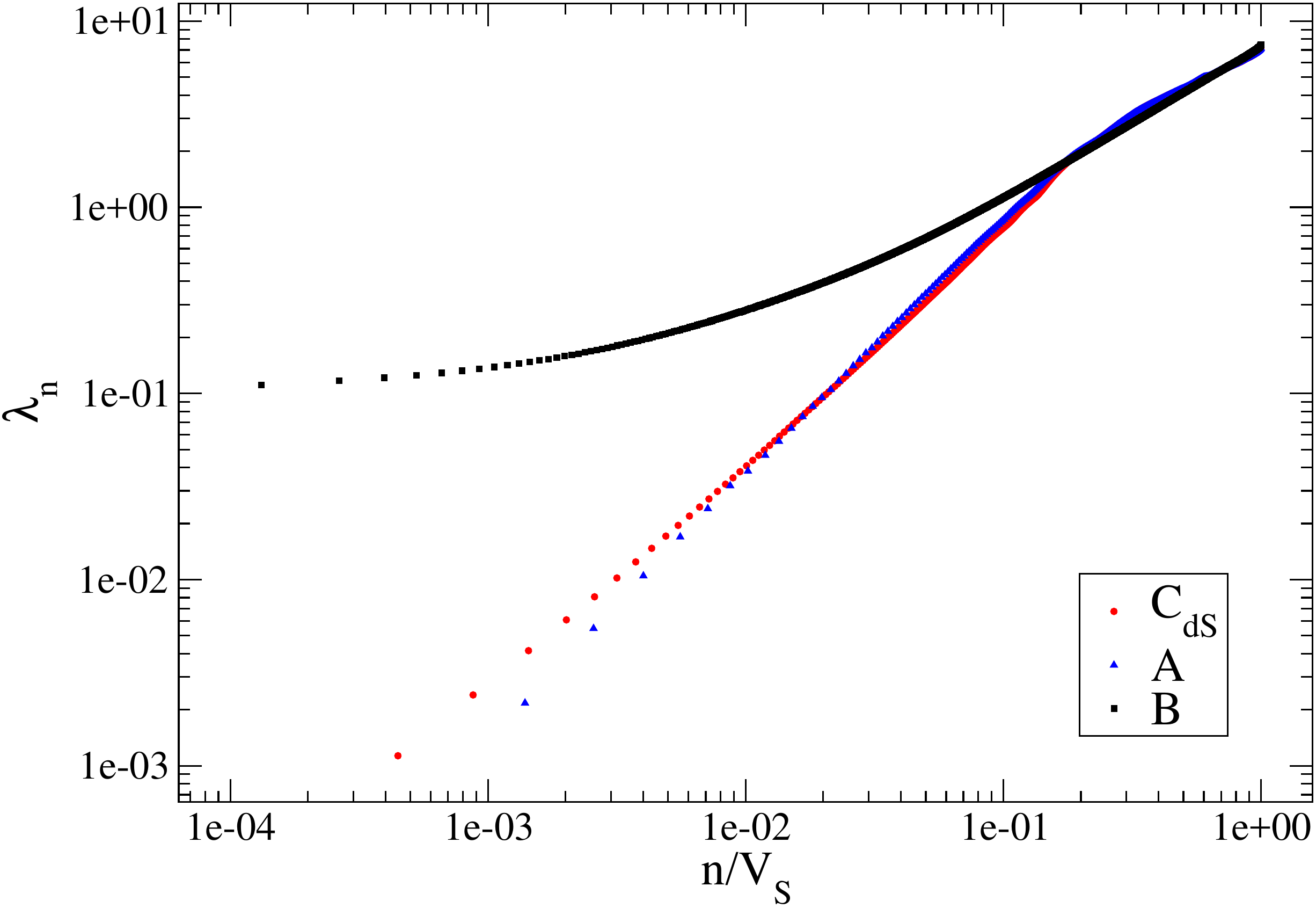}
  	\caption{
  		Averages of $\lambda_n$ versus $n/V_S$ computed in bins of $n/V_S$ with size $2/V_{S,max}$ for slices taken from configurations sampled deep into the $A$, $B$ and $C_{dS}$ phases (simulation points $a$,$b$ and $c$). The volume is fixed to $V_{S,tot}=40k$ for configurations in $A$ and $C_{dS}$ phase, and to $V_{S,tot}=8k$ for configurations in $B$ phase.}
  	\label{fig:averbin_CdS_A_B-lam-vs-k_V}
  \end{figure}

  This kind of scaling is well visible in all phases, as one
  can appreciate by looking at
  Fig.~\ref{fig:averbin_CdS_A_B-lam-vs-k_V}. For convenience, we have
  divided all spatial slices in small volume bins, and then
  averaged $\lambda_n$ for each $n$ over the slices of each bin:
  such averages are reported in the figure against $n/V_S$. 
  average eigenvalues are reported with error bars, which however are too small to be visible.
    
  \begin{figure}
  	\centering
  	\includegraphics[width=1\linewidth]{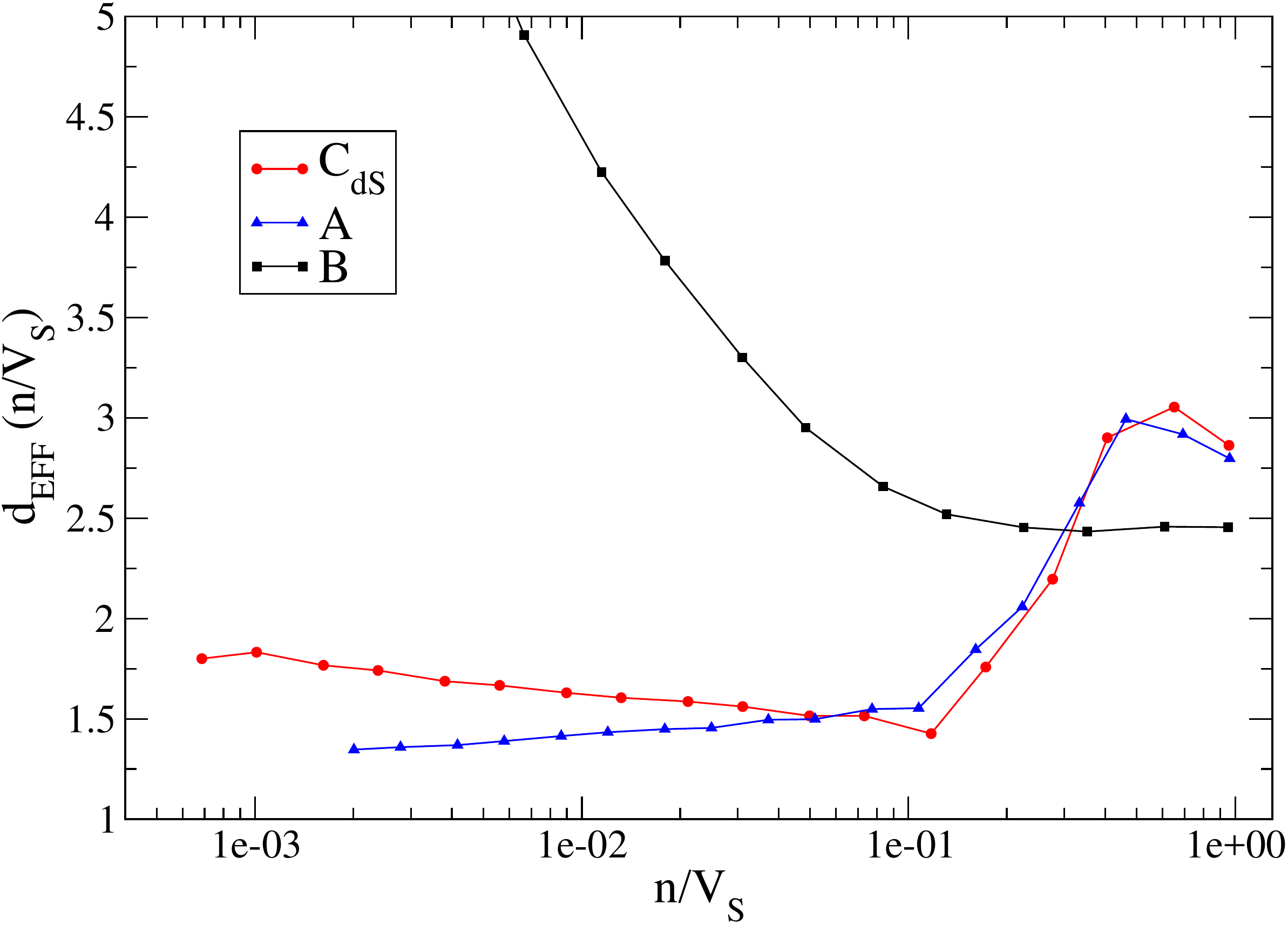}
  	\caption{Running dimension obtained from the logarithmic 
slope $m$ of the curves shown
in Fig.~\ref{fig:averbin_CdS_A_B-lam-vs-k_V} as ${2}/{m}$ 
(see Section~\ref{subsec:toymodel} and Eq.~(\ref{eq:weyl_dimension})), computed over bins of different ranges of $n/V_S$ and for configurations sampled 
in phases $C_{dS}$, $A$ and $B$. The curve associated to the $B$ phase is diverging for $n/V_S \rightarrow 0$ (it is around 30 for 
$n/V_S \sim 10^{-4}$), but part of it has 
been omitted from the plot, to improve the readability of the curves obtained
for the other two phases.}
  	\label{fig:plot_CdS_A_B-Dk_V}
  \end{figure}

  Each phase has its own characteristic profile. The profiles of phases 
  $A$ and $C_{dS}$ are quite similar and differ by tiny deviations:
  in particular, in both cases one has that $\lambda_n \to 0$ 
  as $n/V_S \to 0$, which is an equivalent way to state the absence 
  of a gap in the spectrum. 
  Instead, the profile of phase $B$ is significantly different 
  and characterized by the fact that 
  $\lim_{n/V_S \to \infty} \lambda_n \neq 0$, in agreement with the presence
  of a gap. In 
  Fig.~\ref{fig:averbin_CdS_A_B-lam-vs-k_V} we do not report any data
  regarding the $C_b$ phase, which is discussed separately because of the 
  particular features that we have already illustrated above.
  
  Following the discussion in Section~\ref{subsec:toymodel}, 
  each scaling profile
  can be associated with a running effective dimensionality $d_{EFF}$ of the 
  spatial triangulations at a scale of the order $(n/V_S)^{-1/3}$:
  that can be done by taking
  the logarithmic derivative of $\lambda_n$ with respect to 
  $n/V_S$, see Eq.~\eqref{eq:weyl_dimension}.
  For this reason, in Fig.~\ref{fig:plot_CdS_A_B-Dk_V} we report
  $d_{EFF} = 2\, d \log(n/V_S) / d \log \lambda_n$, which 
  has been computed numerically by taking the average derivative 
  of the profile over small bins of the variable $n/V_S$. 
  
  At very small scales, both the $A$ and the $C_{dS}$ phase are effectively
  3-dimensional. However, going to larger scales (smaller $n/V_S$), 
  the effective dimension decreases, going down to values
  around $d_{EFF} \sim 1.5$, which is approximately the same large scale 
  dimensionality observed by diffusion processes~\cite{cdt_gorlich}. 
  The crossover between the two regimes takes place for 
  $n/V_S$ in the range $0.1  - 0.4$, meaning that typical 
  structures of lower dimensionality develop, with a 
  transverse dimension of the order of just a few tetrahedra.
   
  Actually, the plot of $d_{EFF}$ shows a difference between
  phase $A$ and phase $C_{dS}$, which was not clearly visible before:
  contrary to phase $A$, in phase $C_{dS}$ the effective dimensionality seems
  to slowly grow again as one approaches larger and larger scales.
  This slow grow can be interpreted as a 
  progressive ramification of the lower dimensional
  structures, i.e.~as a hint that it has a fractal-like nature.
  
  The effective dimensionality has a completely different behavior in phase $B$:
  it is smaller than 3 ($d_{EFF} \simeq 2.5$) on small scales, then
  starts growing and diverges at large scales. This is due to the fact
  that $d \log \lambda_n / d \log(n/V_S) \to 0 $ as $n/V_S \to 0$, because
  of the presence of the gap, and, on the other hand, the diverging
  dimensionality can be interpreted in terms of the fact that
  the diameter of the slice grows at most logarithmically with $V_S$.
  Also the low dimensionality observed at small scales can be interpreted 
  in terms of the large connectivity of the associated graphs: each 
  tetrahedron has 4 links to other tetrahedra, some of these
  links are, in some sense, not ``local'', i.e.~they are a shortcut
  to reach directly some otherwise ``far'' tetrahedron;
  then, the probability that a couple of 
  neighbouring tetrahedra are adjacent to a common tetrahedron gets smaller
  and leads to a lower effective dimensionality at short scales.
  \\
  
\begin{figure}[t!]
  	\centering
  	\includegraphics[width=1\columnwidth]{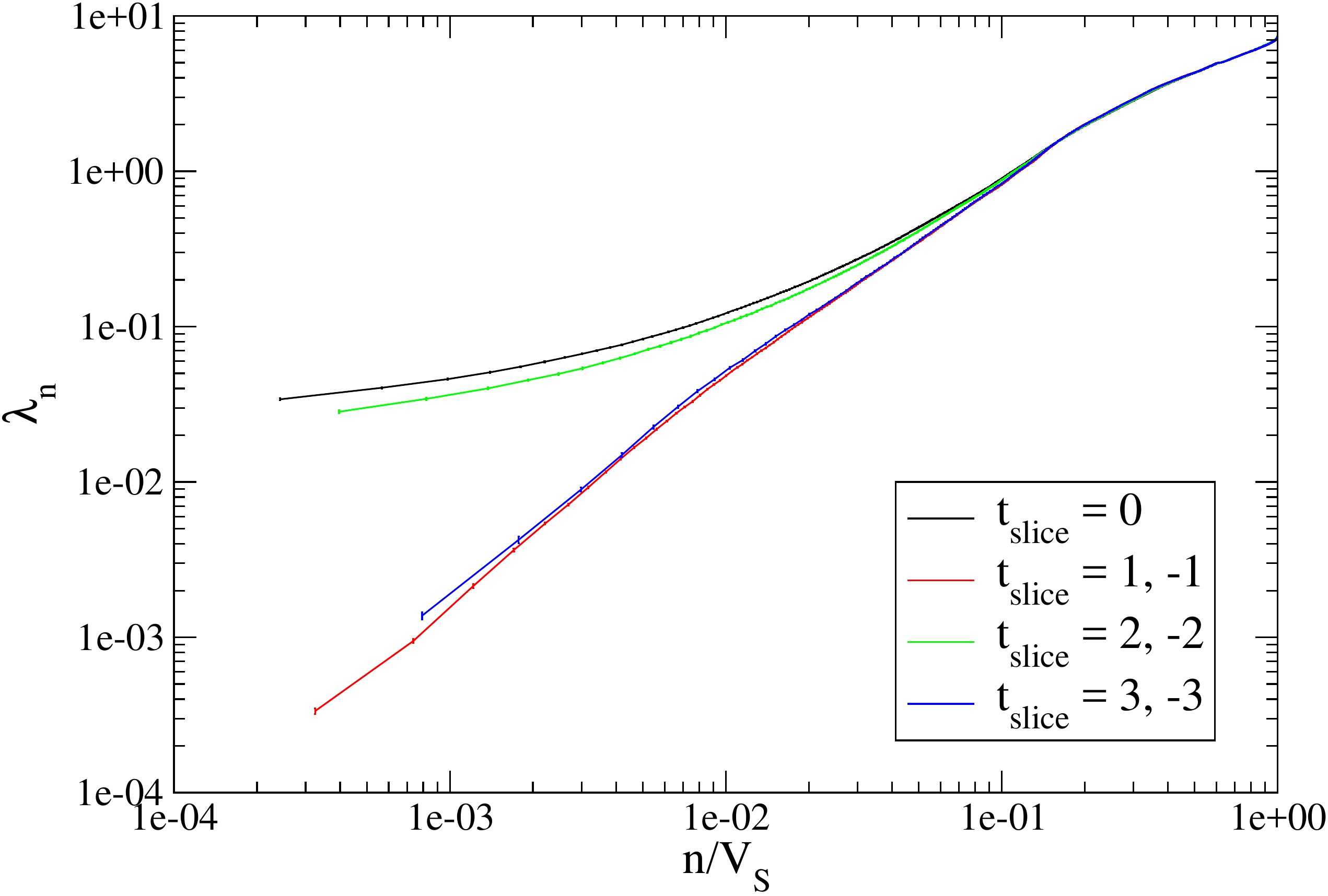}
  	\caption{Averages of $\lambda_n$ versus $n/V_S$ 
  		for slices taken from the bulk ($V_S >1000$) of configurations sampled in the $C_b$ phase 
  		($k_0=2.2$, $\Delta = 0.10$). 
  		The total spatial volume is fixed to $V_{S,tot}=40k$,
  		and the slice times have been relabeled so that the largest 
  		$B$-type slice has $t_{slice} = 0$.}
  	\label{fig:Cb_scaling}
  \end{figure}

  Regarding the properties of the slices found in the bifurcation phase $C_b$,
  on the basis of what we have shown and discussed 
  in Section~\ref{subsec:low_spect}, we have decided to perform a separate 
  analysis for the different classes of spatial slices. 
  In Fig.~\ref{fig:Cb_scaling} we report $\lambda_n$ vs. $n/V_S$ for
  slices according to their relative position with respect to the 
  central largest $B$-type slice (which corresponds to $t_{slice} = 0$).
  The differences between the two classes is clearly visible 
  also from the scaling profiles, which resemble, especially for large scales,
  those found 
  in the $B$ and in the $C_{dS}$ phase for $B$-type and $dS$-type
  slices, respectively. 
  
  However, one striking feature emerges:
  at small scales, in particular for $n/V_S \gtrsim 0.1$, the scaling 
  profiles coincide almost perfectly. We conclude that, at such scales,
  the two classes of slices present strong similarities, despite 
  the completely different large scale behavior. Hints of this fact 
  were already discussed in Section~\ref{subsec:low_spect}. Such similarities 
  are likely induced by the causal structure connecting adjacent
  spatial slices in CDT triangulations.
  
  \begin{figure}[h!!]
  	\centering
  	\includegraphics[width=1\columnwidth]{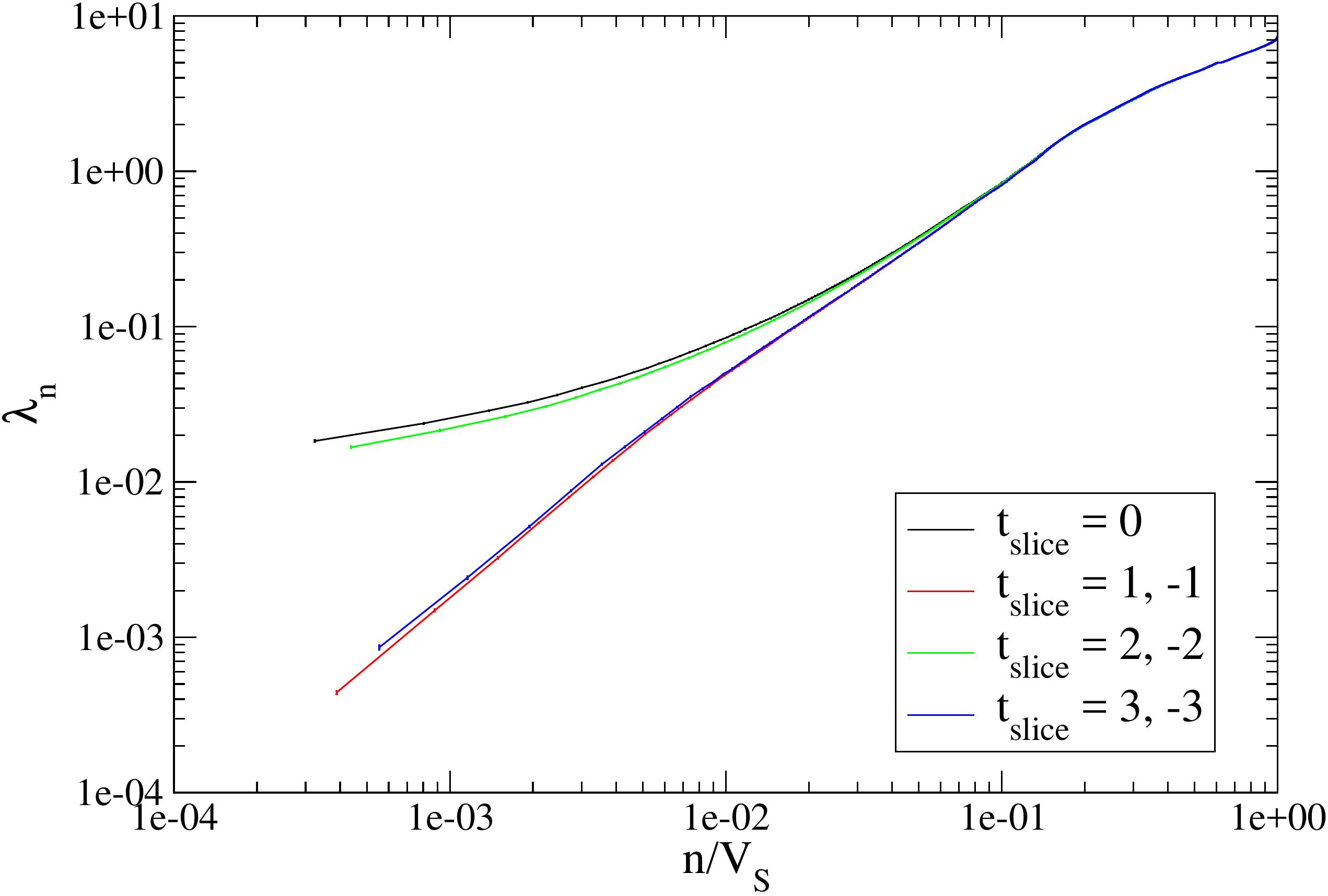}
  	\caption{Averages of $\lambda_n$ versus $n/V_S$ 
  		for slices taken from the bulk ($V_S >1000$) of configurations sampled in the $C_b$ phase 
  		($k_0=2.2$, $\Delta = 0.15$). 
  		The total spatial volume is fixed to $V_{S,tot}=40k$,
  		and the slice times have been relabeled so that the largest 
  		$B$-type slice has $t_{slice} = 0$.}
  	\label{fig:Cb_scaling2}
  \end{figure}
  
  \subsection{Running scales and the search for a continuum limit}

  The analysis of the scaling profiles reported above permits to identify well defined scales, in terms of the parameter $n/V_S$, where something happens, like a change in the effective dimensionality of the system. Such scales are given in units of the elementary lattice spacing of the system, i.e.~the size of a tetrahedron.

  On the other hand, the possible presence of a second order 
  critical point, where a continuum limit can be defined for 
  Quantum Gravity, implies that the lattice spacing should run
  to zero as the bare parameters approach the critical point.
  This running of the lattice spacing should be visible by the 
  corresponding growth of the value, determined in lattice units,
  of some physical scale. This is a standard approach 
  in lattice field theories, where
  one usually considers correlations lengths which are 
  the inverse mass of some physical state.
  
  One of the major challenges in the CDT program is 
  to identify and determine physical scales which could provide 
  such kind of information and thus give evidence that the 
  lattice spacing is indeed running. Promising 
steps in this direction have been already done by means 
 of diffusive processes, where the scale is fixed by the 
diffusion time, both in CDT~\cite{Coumbe:2014noa} and in DT~\cite{dt_syracuse}.
 Here we 
  propose that LB spectra and the observed scaling 
  profiles may be helpful in this direction, 
  and that a careful study of how such profiles 
  change as a function of the bare parameters could 
  provide useful information.
  
  A possible second order point is believed to separate
  the $C_b$ from the $C_{dS}$ phase, therefore it makes
  sense to analyze how the profiles change in both phases
  when moving towards the supposed phase transition, and if 
  the observed changes can be associated to any running scale.
  A growth in the scale associated to some particular feature
  of the scaling profile means that its location moves to 
  smaller values of $n/V_S$.

  As an example, in Fig.~\ref{fig:Cb_scaling2} we report 
  the scaling profiles obtained for
  slices in phase $C_b$ and $\Delta = 0.15$, which is closer
  to the phase boundary with respect to the case
  $\Delta = 0.10$, which has been discussed previously and
  is reported in Fig.~\ref{fig:Cb_scaling}.
  An appreciable difference between the two cases is that
  the region where the profiles of $B$-type and $dS$-type coincide
  is larger (i.e.~extends to smaller $n/V_S$) for 
  $\Delta = 0.15$. From a quantitative point of view, one finds that
  the approximate value of $n/V_S$ where the profiles
  start differing by more than 5\% is around 0.13 for 
  $\Delta = 0.10$ and around 0.074 for $\Delta = 0.15$.
  In other words, there is a scale up to which $B$-type and 
  $dS$-type slices are similar to each other, and such scale
  grows as one approaches the $C_b$-$C_{dS}$ phase transition.
  
\begin{figure}
  	\centering
  	\includegraphics[width=1\linewidth]{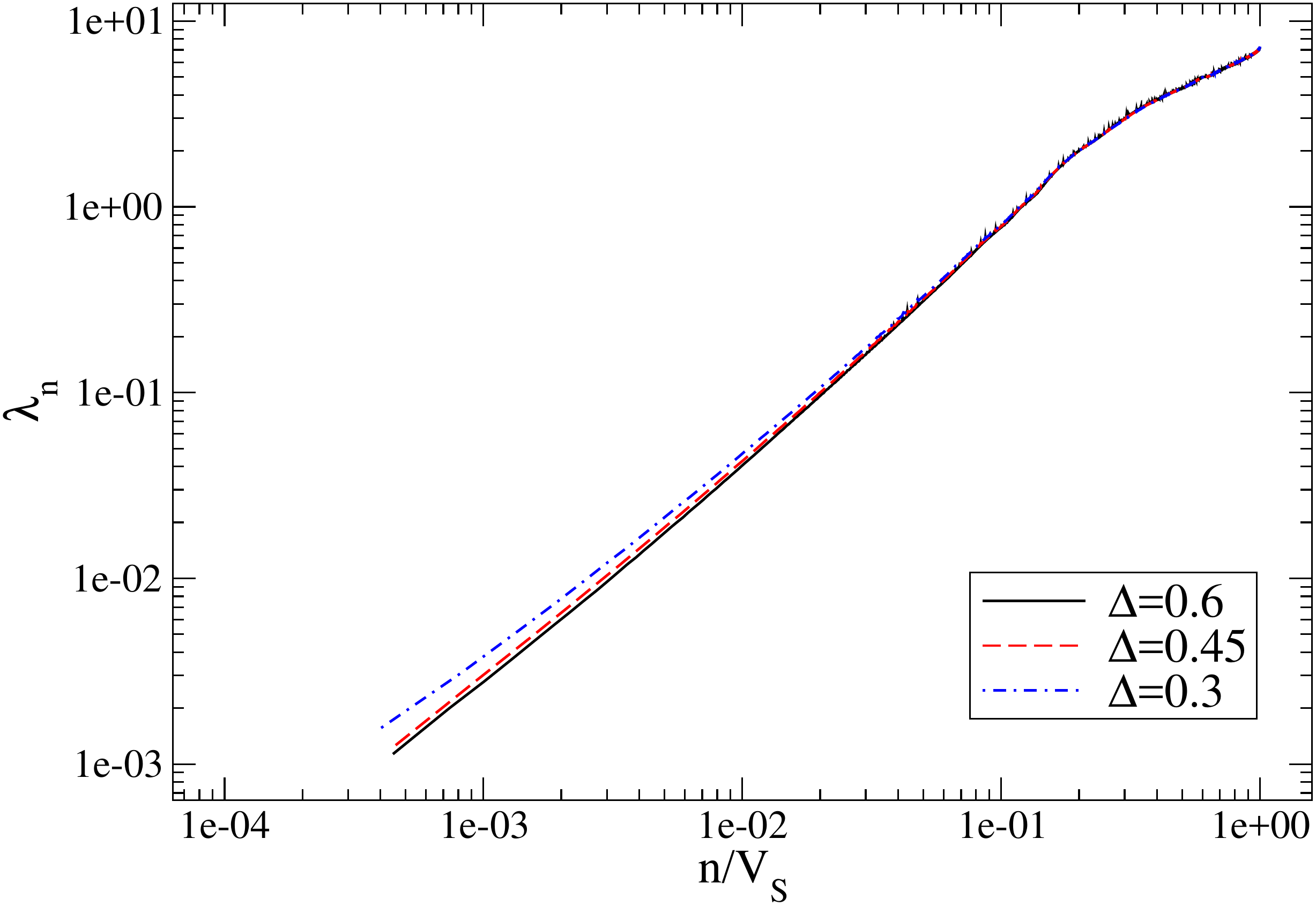}
  	\caption{
  		Averages of $\lambda_n$ versus $n/V_S$, 
computed in bins of $n/V_S$ with size $2/V_{S,max}$, for slices taken 
from configurations sampled in the $C_{dS}$ phase, with $k_0=2.2$ and different values of $\Delta$. The total spatial volume of each configuration is $V_{S,tot}=40k$. 
  	}
  	\label{fig:averbin_CdS123-lam-vs-k_V}
  \end{figure}

  In a similar way, one can look at how the scaling profiles found 
  in the $C_{dS}$ phase change as one approaches the phase
  transition from the other side. Such scaling profiles are reported 
  in Fig.~\ref{fig:averbin_CdS123-lam-vs-k_V}. The short-scale region,
  and in particular the point where the effective dimension starts
  changing, seems not sensible to the change of $\Delta$. However 
  the small $n/V_S$ (large-scale) region changes, with the profile
  undergoing an overall bending towards the left: notice that 
  this implies a change in the effective dimensionality observed 
  at the largest scales, which indeed, for $\Delta = 0.3$, 
is $d_{EFF}\gtrsim 2$.
  
  Finally, as we have already stressed above, the gap itself,
  	which for $B$-type slices seems to approach zero 
  	as one gets closer to the $C_b$-$C_{dS}$ phase transition
  	(see Fig.~\ref{fig:lambda1_op-scatt})
  	could be interpreted in terms of a diverging correlation length
  	if the behavior is proved to be continuous.
  
  The reported examples are only illustrative of the fact that
  the LB spectrum can provide useful scales which could
   give information on the nature
  of a possible continuum limit.
Such program should be carried
  on more systematically by future studies, in particular by approaching
  the $C_b$-$C_{dS}$ phase transition more closely.

  \subsection{Fine structure of the full spectrum}\label{subsec:full_spectrum}

\begin{figure}[ht]
\centering
\includegraphics[width=1\linewidth]{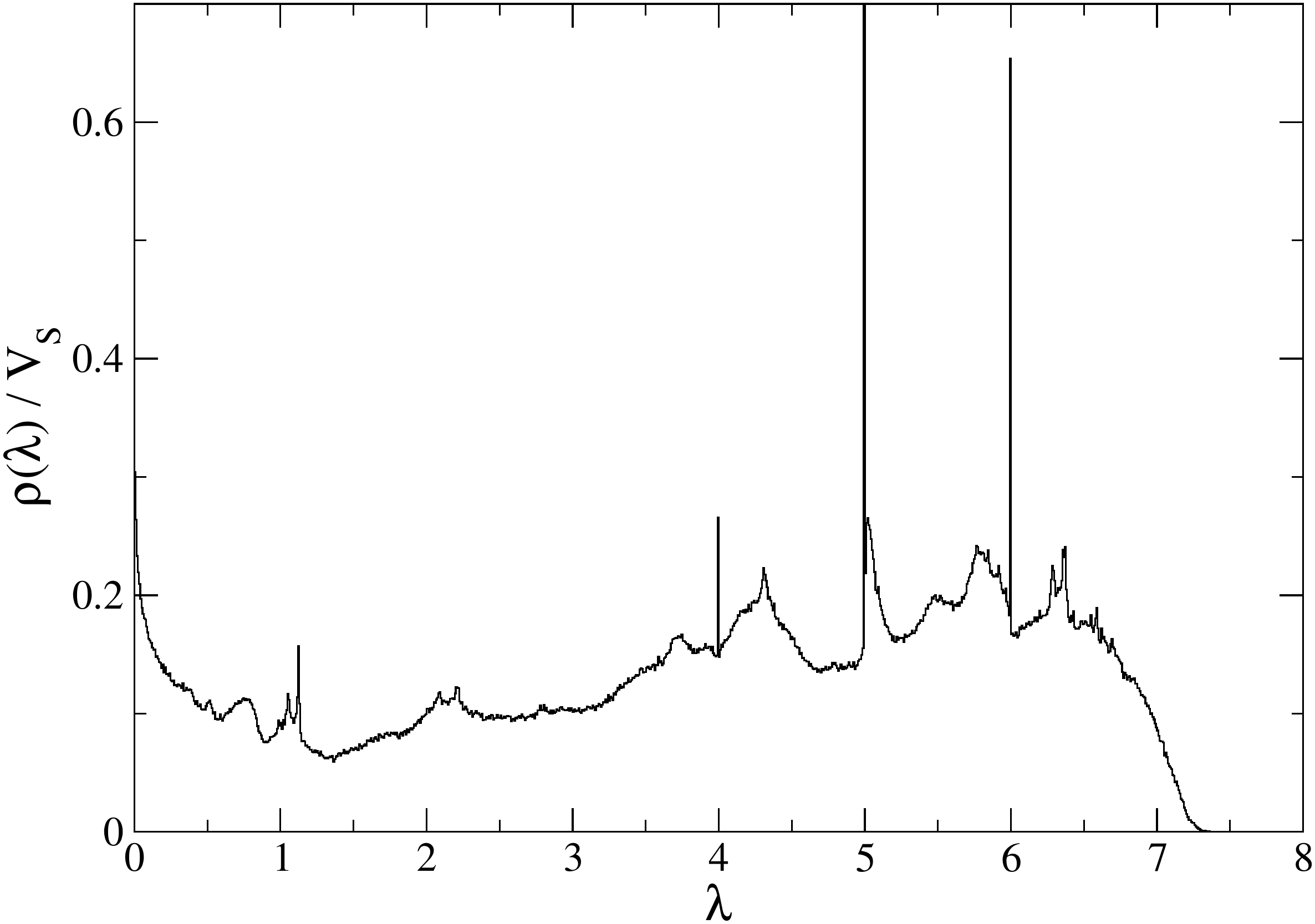}
\caption{Normalized distribution of all the eigenvalues for slices with volume in the range $V_S\in [2000,2500]$ for configurations deep into the $C_{dS}$ phase (simulation point $c$), and with total spatial volume $V_{S,tot}=40k$.\label{fig:fullspectrum_CdS_histo}}
\vspace*{\floatsep}
\includegraphics[width=1\linewidth]{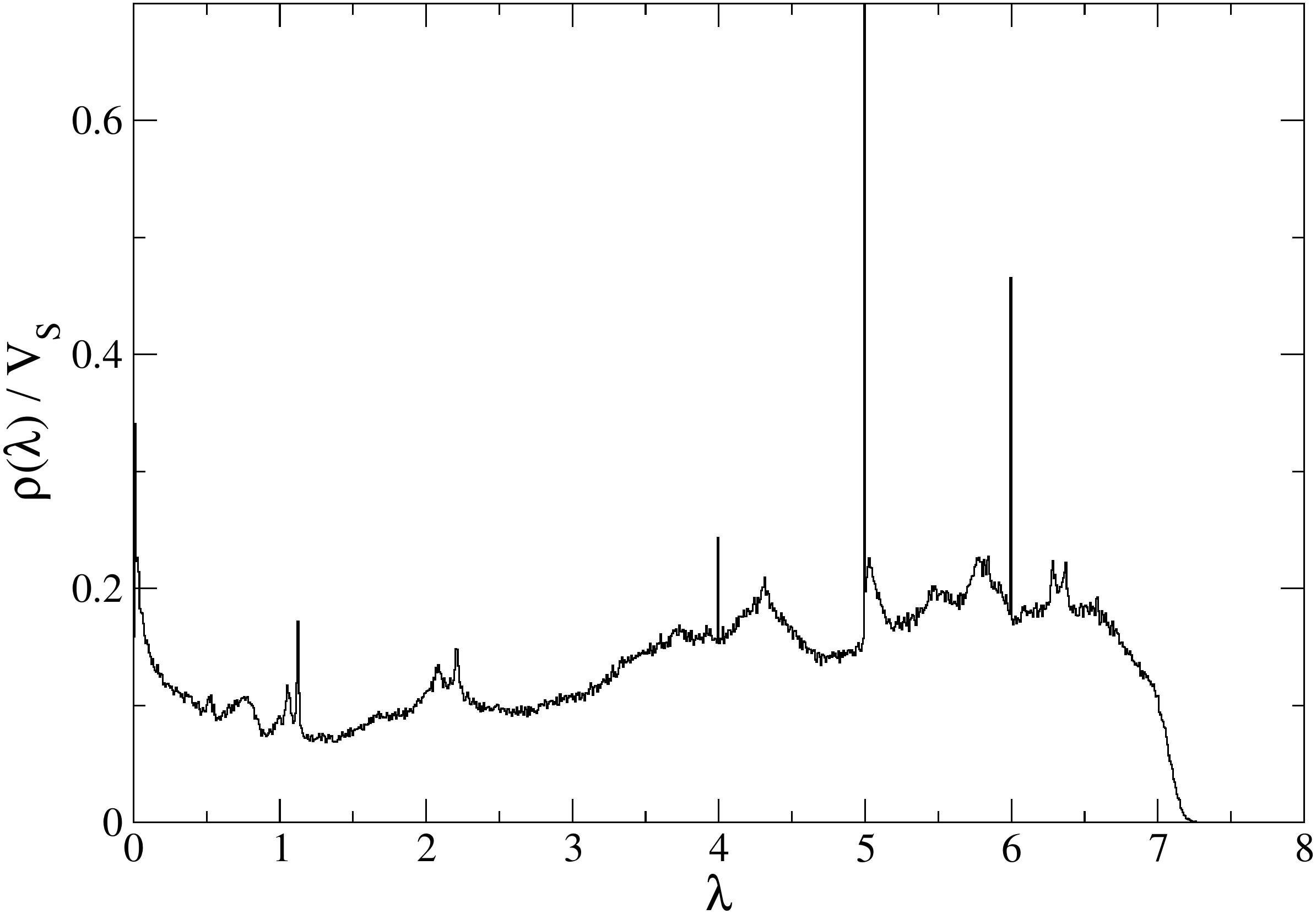}
\caption{Normalized distribution of all the eigenvalues for slices with volume in the range $V_S\in [200,400]$ for configurations deep into the $A$ phase (simulation point $a$), and with total spatial volume $V_{S,tot}=40k$.\label{fig:fullspectrum_A_histo}}
\end{figure}

\begin{figure}[h!]
  	\centering
  	\includegraphics[width=1\linewidth]{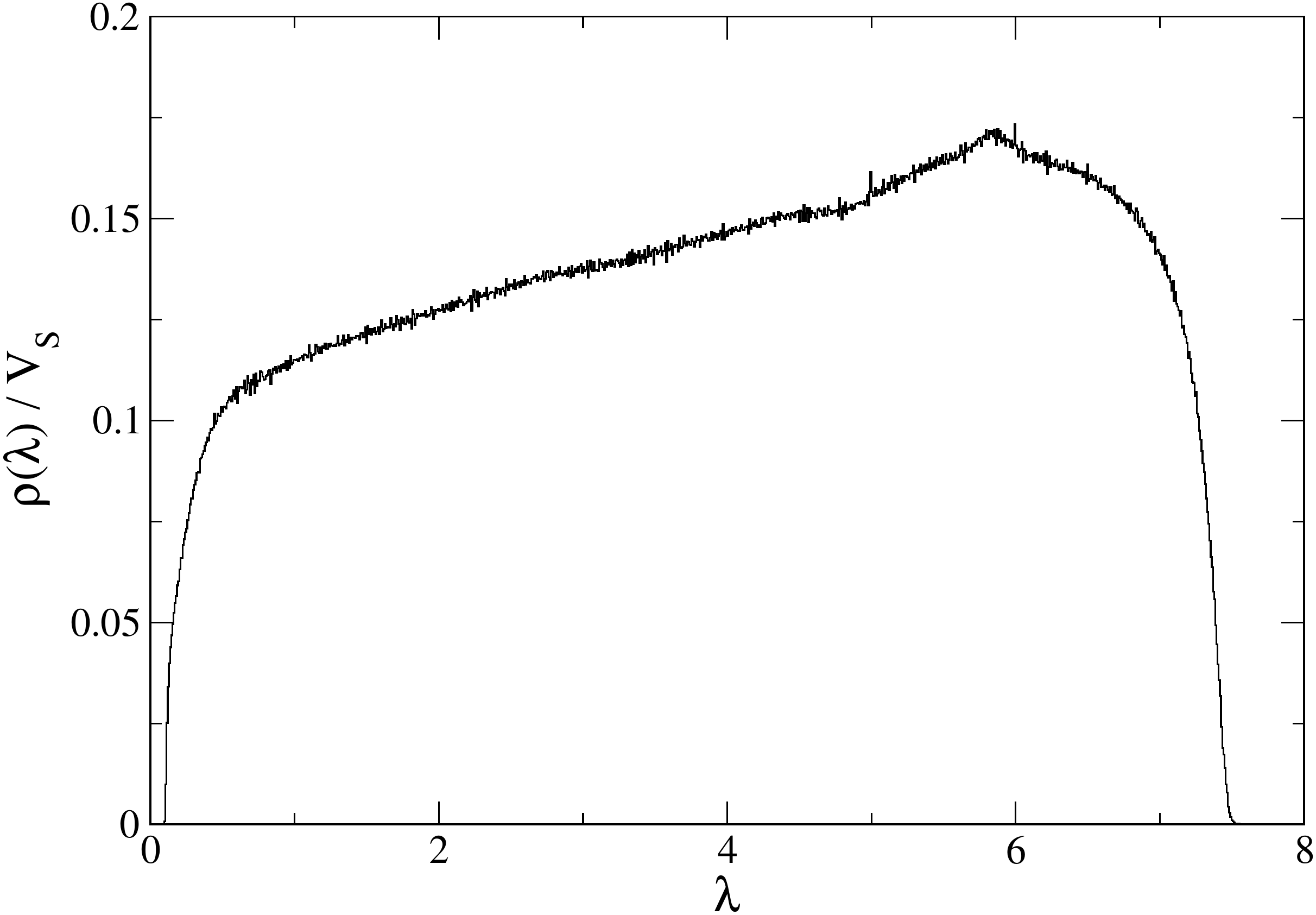}
  	\caption{
  		Normalized distribution of all the eigenvalues of the maximal slices for configurations deep into the $B$ phase (simulation point $b$), and with spatial volume about $V_S \simeq 8k$.}
  	\label{fig:fullspectrum_B_histo}
  \end{figure}

  In this section we will show some details regarding
  the full distribution of eigenvalues (i.e.~over the whole spectrum) 
  in the different phases. 
  Figs.~\ref{fig:fullspectrum_CdS_histo}, \ref{fig:fullspectrum_A_histo} and \ref{fig:fullspectrum_B_histo} show the normalized distribution
  of eigenvalues for spatial slices with volumes in selected ranges, and for 
  simulations performed deep into the phases $C_{dS}$, $A$ and $B$ respectively.

  The $A$ and the $C_{dS}$ phase present a detailed non-trivial
  fine structure which is very similar. Even if we are not interested,
  at least in the present context, to provide a detailed 
  interpretation of the full spectrum, we notice that such fine structure
  is mostly relative to eigenvalues which are of order 1 or larger, hence 
  associated to typically small scales; this is confirmed by the fact that,
  contrary to the low part of the spectrum, such fine structure is almost
  left invariant by changing the volume of the slice.
  For instance, it can be noticed that the distributions 
  are sharply peaked around 
  the integer values $\lambda=4,5,6$; indeed, by inspecting the spectra of 
  single configurations and the associated eigenvectors, 
  we observed that these integer eigenvalues often 
  occur with high multiplicity and can be associated to the presence
  of recurrent regular short-scale structures and to very localized
  eigenvectors.
  
  The normalized distribution in the case of configurations in the $B$ phase 
  does not show particular features, other than the already discussed presence 
  of a spectral gap. The distribution looks in general more regular in this 
  phase, even if some of the peaks around integer values are still 
  present, but much reduced in amplitude.

 \subsection{Visualization of spatial slices}\label{subsec:visualization}

  We have seen how each spatial slice of the triangulations can
  be associated to a graph with non-trivial properties, i.e.~what
  is usually called a {\em complex network}. 
  There are different methods to visualize a complex network,
  some of them already considered in previous studies 
  (see, e.g., Ref.~\cite{dt_syracuse}),
  here we will briefly discuss only two of them: \emph{Laplace embedding}~\cite{LB_embedding} and \emph{spring embedding}~\cite{force-directed_embeddings,spring_embedding}. The former makes use of the eigenvectors associated to the smallest eigenvalues, which are already computed by solving the eigenvalue problem, while the latter is based on 
  a mapping of the graph to a system of points connected
  by springs: as we are going to discuss, the two methods
  are strictly related, however spring embedding proves more useful
  to give an intuitive picture of the short-scale structures. 
  
  The underlying 
  idea, common to both methods, is to represent any graph $G=(V,E)$ in a 
  $m$-dimensional Euclidean space by finding a set of $m$ independent 
  functions $\{\phi_n(v_i)\}_{n=1}^{m}$ which act as coordinates for each 
  vertex $v_i\in V$, in such a way that vertices with smaller graph distance 
  have coordinates with values as closer as possible. The ``closeness''
  can be defined in many ways, 
  consisting in solving different optimization problems,
  and that makes the two methods different. 
  We will use the notation $\vec{\phi}_n \equiv (\phi_n(v_i))_{i=1}^{|V|}$ for each $n=1,\dots, m$.

  \subsubsection{Laplace embedding}
  The optimization problem for Laplace embedding~\cite{LB_embedding} consists in minimizing the following functional of the coordinate functions $\{\phi_n\}$:
  \begin{align}
  \mathcal{E}_{LB}[\phi] &\equiv \frac{1}{2} \sum\limits_{n=1}^{m} \sum\limits_{(v_i,v_j)\in E} \Big( \phi_n(v_i)- \phi_n(v_j) \Big)^2 \nonumber \\
  &= \frac{1}{2} \sum\limits_{n=1}^{m} \vec{\phi}_n^T L \vec{\phi}_n \, .
  \end{align}
  subject to the constraints $\vec{\phi}_n \cdot \vec{\phi}_k = \delta_{n,k}$ and $\vec{\phi}_n \cdot \vec{1} = 0$ for each $n,p=1,\dots,m$, where
  $\vec{1}$ is the uniform vector with unit coordinates and $L$ 
  is the matrix representation of the LB operator.
  It is straightforward to prove that a solution to this constrained optimization problem is given by the set of the first $m$ eigenvectors $\{\vec{e}_n\}_{n=1}^{m}$ of the Laplace--Beltrami matrix, where we excluded the $0$-th mode $\vec{e}_0 = \frac{1}{\sqrt{|V|}}\vec{1}$ (second constraint) and ordered eigenvectors without multiplicity (i.e.~$\vec{e}_n$ is associated to $\lambda_n$ and $\lambda_n \leq \lambda_{n+1}$).

\begin{figure}[t!]
  	\centering
  	\includegraphics*[width=1.1\linewidth,trim=100 0 0 0]{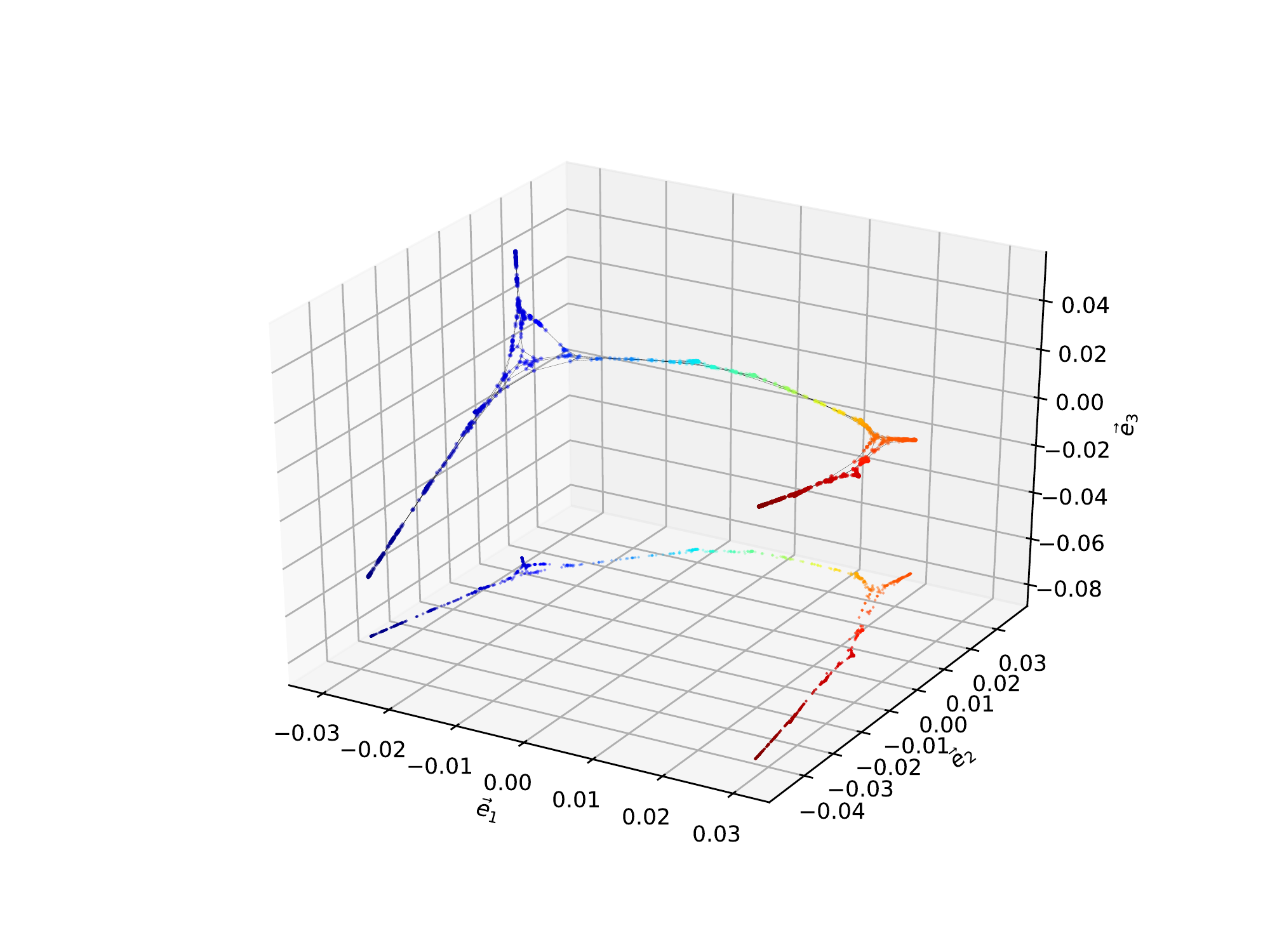}
  	\caption{Laplace embedding in $3$ dimensions for the graph associated to a typical slice in the $C_{dS}$ phase (simulation point $c$) and with volume $V_S\simeq1500$. Here the color identifies the values that takes the first eigenvector $\vec{e}_1$ on each vertex: blue is negative, green is zero and red is positive.
  		A projection of the $3$-dimensional figure is shown on the $xy$ plane.}
  	\label{fig:LB_bf51_3D}
  \end{figure}

  For example, the coordinates associated to each vertex $v\in V$ in a $3$-dimensional Laplace embedding are the values of the first $3$ eigenvectors on that vertex, that is $v \mapsto (e_1(v),e_2(v),e_3(v))\in \mathbb{R}^3$. Fig.~\ref{fig:LB_bf51_3D} shows the $3$-dimensional Laplace embedding of a typical slice in the bulk of a configuration deep in the $C_{dS}$ phase (simulation point $c$). The geometry seems to be made up of filamentous structures, but that really means that the first $3$ eigenvectors, describing the slowest modes of diffusion, are not capable of describing short scale structures inside the filaments. However, they efficiently describe the largest scale geometry, which in the $C_{dS}$ case is 
  non-trivial and unexpected.

  \begin{figure}[t!]
  	\centering
  	\includegraphics*[width=1.2\linewidth,trim=150 0 0 0]{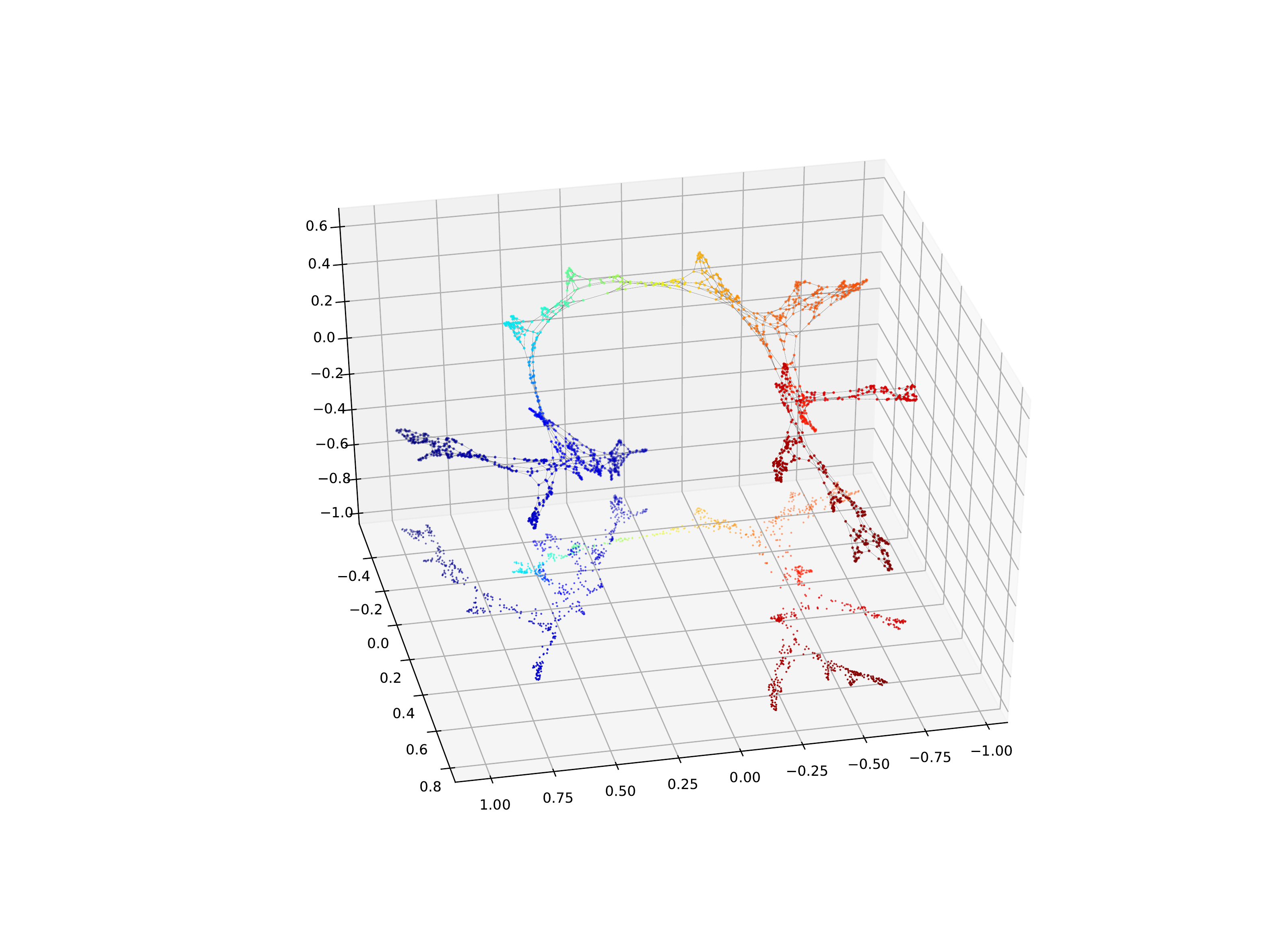}
  	\caption{Spring embedding in $3$ dimensions for the graph associated to a typical slice in the $C_{dS}$ phase; the slice is the same as in Fig.~\ref{fig:LB_bf51_3D}. The rest length has been fixed to  $l_0=0.02$. 
Also in this the color identifies the values that the first eigenvector $\vec{e}_1$ takes on each vertex, see Fig.~\ref{fig:LB_bf51_3D}.
  		A projection of the $3$-dimensional figure is shown on the $xy$ plane.}
  	\label{fig:spring_bf51_3D_v2}
  \end{figure}

  \subsubsection{Spring embedding}
  
  The optimization problem that has to be solved for spring embedding of an unweighted undirected graph $G=(V,E)$ consists in the energy minimization of a system of ideal springs with fixed rest length 
  $l_0$ and  embedded in $\mathbb{R}^m$, with extrema connected in the same way as the links of the abstract graph $G$~\cite{spring_embedding}. Having assigned coordinates $\{\phi_n(v_i)\}_{n=1}^{m}$ to each abstract vertex of the graph $v_i\in V$, the potential energy of the system is defined as:
  \begin{equation}
  \mathcal{E}_{S}[\phi] = \frac{1}{2} \hspace{-0.15cm}
  \sum\limits_{(v_i,v_j)\in E} \hspace{-0.15cm}
  \left( \hspace{-0.1cm}
  l_0-\sqrt{\sum\limits_{n=1}^{m} (\phi_n(v_i)- \phi_n(v_j))^2}\right)^2 
  \end{equation}
  In the limit $l_0\rightarrow 0$, the functional $\mathcal{E}_{S}$ becomes equal to $\mathcal{E}_{LB}$ but with 
  no constraint, so that the solution would collapse to the trivial solution,
  bringing all vertices to the same point, in this limit. On the other hand,
  for $l_0 > 0$, the springs will push vertexes apart from each other and 
  help resolving even the shortest-scale structures, which are not visible
  with Laplace embedding.
  
  The simplest algorithm to find a (local) minimum is to initialize the coordinates of each vertex to a random value, and then relax the system of springs by performing a gradient descent. Fig.~\ref{fig:spring_bf51_3D_v2} we shows the spring embedding of the same slice represented by Laplace embedding in Fig.~\ref{fig:LB_bf51_3D}. The large scale structure is well represented by both methods, but spring embedding permits 
  to better discern short-scale structures at the finest level.

  Such representations of the spatial slices are illuminating to 
  understand the properties of the LB spectrum for $C_{dS}$ slices.
  The slices are extended objects, i.e.~one finds vertexes which are far
  apart from each other, implying the existence of slow diffusion modes and a continuum
  of quasi-zero eigenvalues for large $V_S$. 
  On the other hand, the large scale structure is made of 
  lower-dimensional substructures, which have a typical transverse size
  of the order of a few vertexes, and which often branch, making the 
  overall spectral dimension (i.e.~the diffusion rate) fractional
  at large scales.

\begin{figure}
  	\centering
  	\includegraphics*[width=1.1\linewidth,trim=100 0 0 0]{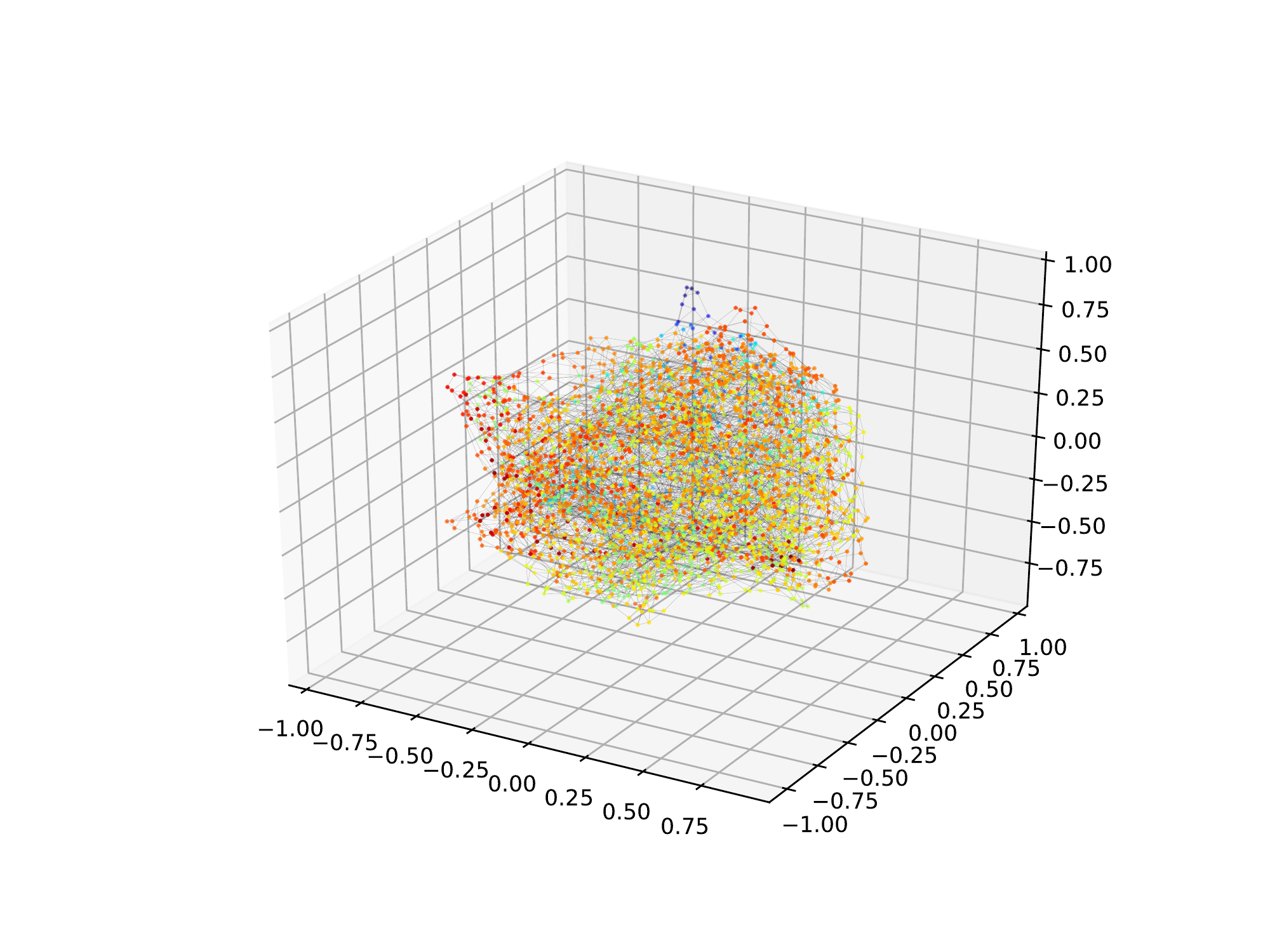}
  	\caption{Spring embedding ($l_0=0.015$) in $3$ dimensions for the graph associated to a typical slice deep in the $B$ phase (simulation point $b$) and with volume $V_S\simeq4000$. 
Also in this the color identifies the values that the first eigenvector $\vec{e}_1$ takes on each vertex, see Fig.~\ref{fig:LB_bf51_3D}.}
  	\label{fig:spring_B10k_it500_3D_v3}
  \end{figure}

  For comparison, Fig.~\ref{fig:spring_B10k_it500_3D_v3} shows the spring embedding of a typical slice in $B$ phase. The high connectivity of the graph, which is clearly visible
  from the figure, does not permit the development of extended large
  scale structures, so that diffusion modes maintain always fast 
  and a finite gap remains even in the $V_S \to \infty$ limit.

 \section{Discussion and Conclusions}\label{sec:conclusions}

In this work we have investigated the properties of the
different phases of CDT that can be inferred from an analysis
of the spectrum of the Laplace-Beltrami operator computed on the 
triangulations. The present exploratory study has been limited 
to the properties of spatial slices: those can be associated
to regular graphs where each vertex is linked to 4 other 
vertices. Let us summarize our main results 
and further discuss them:
\\

{\em i)\,} We have shown that the different phases can be characterized
according to the presence or absence of a gap in the spectrum,
which therefore can be considered as new order parameter
for the phase diagram of CDT. In particular,
a gap is found in the $B$ phase,
while for the 
$A$ and the $C_{dS}$ phases one finds a non-zero density 
of eigenvalues around $\lambda = 0$ in the thermodynamical
(large spatial volume $V_S$) limit. The $C_b$ phase, instead, shows 
the alternance of spatial slices of both types (gapped and 
non-gapped): that better characterizes the nature of the alternating 
structures already found in previous works~\cite{cdt_newhightrans,cdt_charnewphase}, which for this reason
we have called $B$-type and $dS$-type slices.

The presence or absence of a gap
in the spectrum is a characteristic which distinguishes different
phases in many different fields of physics: think for instance of 
Quantum Chromodynamics, where the absence/presence of a gap in the 
spectrum of the Dirac operator characterizes the phases with 
spontaneously broken/unbroken chiral symmetry.

In this context, the presence of a gap tells us that the spatial slices
are associated to expander graphs, characterized by a high
connectivity. That can be interpreted geometrically as a Universe 
with an infinite dimensionality at large scales,
with a diameter which grows at most logarithmically in the 
thermodynamical limit; a small diameter in the 
phases with a gap is consistent with the findings of previous studies
and is supported by a direct computation (see Fig.~\ref{fig:diam_CdS_B_ecc-vs-V}).
On the contrary, the closing of the gap
can be interpreted as the emergence of a Universe with a 
standard finite dimensionality at large scales.
It is interesting to notice that the value of the gap
which is found seems to change continuously as one moves
from the $B$ to the $C_b$ phase, and approaches zero
as the $C_{dS}$ phase is approached.
\\

{\em ii)\,} We have shown that the spectrum can be characterized
by a well defined scaling profile: the 
$n$-th eigenvalue, $\lambda_n$, is a function of just the scaling
variable $n/V_S$.
The profile is different for each phase and characterizes it; moreover, 
from the profile one can deduce information on the effective dimensionality $d_{EFF}$ 
of the system at different scales, which generalize a similar kind of information gained by diffusion processes.

The $C_{dS}$ and the $A$ phase share a similar profile,
corresponding to $d_{EFF} \simeq 3 $ at short scales, which then drops to
$d_{EFF} \simeq 1.5 $ for $n/V_S \lesssim 0.1$. At larger scales,
the two phases show a different behavior, with $d_{EFF}$ which 
keeps decreasing as $n/V_S$ decreases in the $A$ case, while 
in the $C_{dS}$ phase it starts growing again at large scales.
Slices in the  $B$ phase, instead, show an effective dimensionality which,
in agreement with their high connectivity, seems to diverge in the large scale
limit.

An interesting feature has been found for the two different and 
alternating (in Euclidean time) classes
of spatial slices in the $C_b$ phase: despite the 
different overall structure, they
share an identical profile at small length scales, which 
is likely induced by the causality condition imposed
on triangulations and is therefore an essential property of CDT.
The profiles remain identical up to characteristic length
scale above which they start to diverge, as expected
since one class presents a gap and the other does not.
\\

{\em iii)\,} We have proposed that the scaling profiles might be used
to identify particular length scales which change as a function
of the bare parameters, and thus could serve as possible probes
of the running to the continuum limit, if any. Among those, we have
found of particular interest the characteristic length scale up to which
the alternating slices found in the $C_b$ phase share the same profile:
we have seen that such length grows as one approaches the boundary
with the $C_{dS}$ phase. On the other side of the boundary,
also the profiles of the slices in $C_{dS}$ phase show a modification
at large scales as the $C_b$ phase is approached, leading in
particular to a growing effective dimensionality. 

Along these lines,
one could conjecture that, if a second order critical point is 
really found between the two phases, at such a point the 
different profiles found in the $C_b$ phase could merge
at all scales and coincide with the profile
from the $C_{dS}$ phase. Such a critical point would 
also been characterized by the vanishing of the 
gap for the $B$-type slices of the $C_b$ phase. Moreover,
it would be interesting to test what the effective dimensionality
found at large scales would be at the critical point: is it possible that,
just on the critical point where a continuum limit can be defined,
the effective dimensionality of spatial slices goes back 
to $D = 3$ at all physical scales? 
\\

The present work can be continued along many directions.
First of all, the region around the transition between the 
$C_b$ and the $C_{dS}$ phase should be studied in much more
detail than what done in the present exploratory work, to see
if some of the conjectures that we have made above can be put 
on a more solid basis.
In addition, 
a careful study of the critical behavior around the transition of the 
spectral gap, which is the new order parameter introduced in this 
study, could provide information about the universality class
to which the continuum limit, if any, belongs.
Of course, it could well be that one finds a first order transition,
i.e.~a sudden jump in the gap and in other properties, 
but then one should perform simulations for lines corresponding 
to different $k_0$ to see if the first order terminates at some
critical endpoint.
 
We have not considered yet the information which can be gained
by inspecting the eigenvectors of the LB operator, that will
be done in a forthcoming study. In particular, it will be interesting
to consider and analyze their localization/delocalization 
properties, in a way similar to what has been done in similar
studies for the spectrum of the Dirac operator in QCD~\cite{qcd_anderson_multifractal,qcd_anderson_eigenmodes}.

It will be interesting to extend the study of the spectrum
to the full triangulations, i.e.~not just for spatial slices.
That will require some implementative effort: unlike spatial tetrahedra (which are all identical), 
pentachorons can have edges with different Euclidean lengths, and therefore a regular graph representation 
does not describe the geometry faithfully. Nevertheless, the Laplace--Beltrami operator for general 
triangulated manifolds would have 
a well defined representation in the formalism of Finite Elements Method, 
as discussed and applied for example in Refs.~\cite{reuter_cad,reuter_dna}.

Finally, it would be interesting to apply spectral methods
also to other implementations of dynamical triangulations, like the
standard Euclidean Dynamical Triangulations (DT) where
no causality condition is imposed. The implementation
in this case would be straightforward, as for the spatial slices
of CDT, i.e.~given in terms of regular indirected graphs.
We plan to address the issues listed above in the next future.

\acknowledgements
 
We thank C.~Bonati and M.~Campostrini
for useful discussions, and 
 A.~G\"orlich for giving us access
to a version of CDT code which has served 
for a comparison with our own code.
Numerical simulations have been performed at the Scientific Computing
Center at INFN-PISA. 

 \appendix
 \section{Heat-kernel expansion and spectral dimension}\label{sec:heatkernel}
 Here we describe how to compute the spectral dimension of a graph $G$ from the spectrum of its LB matrix using the heat-kernel expansion. Later we will apply the definition to slices in $C_{dS}$ phase, making a comparison between the standard definition of spectral dimension (i.e.~via diffusion processes), and the one obtained by the spectrum.

 Let us consider the fundamental solution to the diffusion equation on a $k$-regular connected graph $G$ with LB matrix $L$:
 \begin{equation}\label{eq:HKernel_eq}
 \begin{cases}
 \partial_t K_{v,v_0}(t) = -\frac{1}{k} \sum\limits_{v' \in V} L_{v,v'} K_{v',v_0}(t) \\
 K_{v,v_0}(0) = \delta_{v,v_0} \, .
 \end{cases}
 \end{equation}
 Discretizing time with unit steps $\Delta t = 1$~\cite{cdt_spectdim}, Eq.~\eqref{eq:HKernel_eq} becomes the equation for random walk on the graph, where $K_{v,v_0}(\tau)$ is the probability that a random walker starting from the vertex $v_0$ at time $t=0$ is found at the vertex $v$ at time $t=\tau$.
 The \emph{return probability} $Z_{v_0}(t) = K_{v_0,v_0}(t)$ is the probability that a random walker comes back to the starting vertex $v_0$ after $t$ steps. Averaging over all starting vertices, the return probability reduces to $Z(t)=\tr K (t)$, the heat-kernel trace. In practice, the diffusion is performed only for a random subset of starting vertices, from which the return probability is then estimated as an average over explicit diffusion processes.

However, $Z(t)$ can also be computed using the spectrum of the LB matrix using the \emph{heat-kernel} expansion for the solution to Eq.~\eqref{eq:HKernel_eq}:
 \begin{equation}\label{eq:HKernel}
 K_{v,w}(t) = \frac{1}{|V|} \sum\limits_{n=0}^{|V|-1} e_{n}(v) e_{n}(w) e^{-\lambda_n t / k} \, ,
 \end{equation}
 where $\{\lambda_n\}$ and $\{\vec{e}_n\}$ are the eigenvalues and associated eigenvectors of the LB matrix of the graph $G$. Notice that the terms in Eq.~\eqref{eq:HKernel} corresponding to larger eigenvalues are more suppressed for increasing times than terms corresponding to smaller ones. In particular, for times $t \gg k/\lambda_1$, the only surviving term is given by the $0$-th eigenvalue, and the probability distribution tends to be uniformly distributed amongst all vertices: $\lim_{t \rightarrow +\infty} K_{v,v_0}(t) = \frac{1}{|V|} \;\forall v,v_0 \in V$ (assuming a single connected component).

\begin{figure}
\centering
\includegraphics[width=1.0\linewidth]{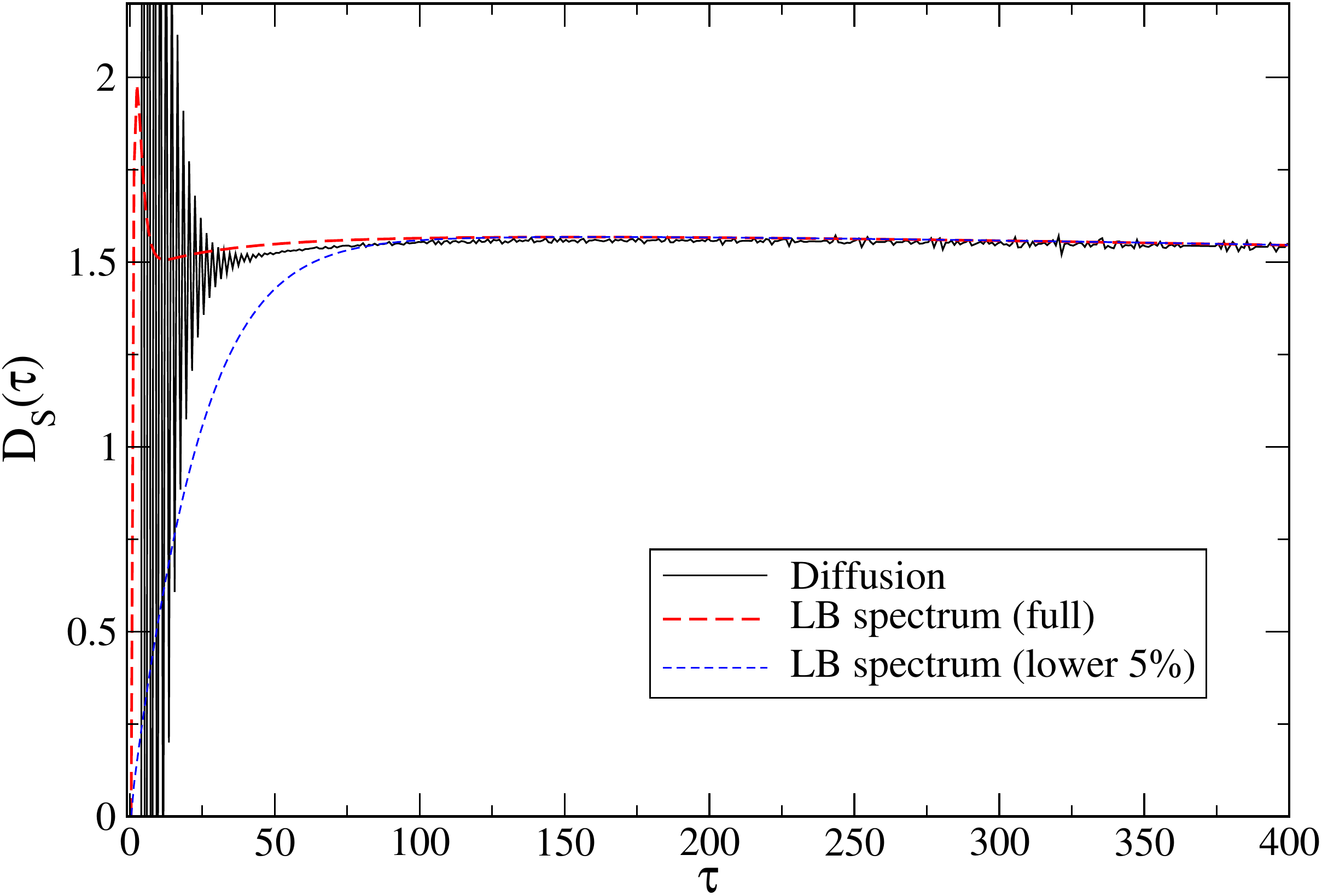}
\caption{Estimates of the running spectral dimension (see Eq.~\eqref{eq:sptdim_formula}) obtained either via diffusion processes (continuous line), or using Eq.~\eqref{eq:Zmanif} (dashed lines) with the full spectrum or only the lowest $5\%$ part of it, for slices in the volume range $2000 - 2200$ taken 
from configurations sampled in the $C_{dS}$ phase (simulation point $c$) with total spatial volume $V_{S,tot}=40k$.}
\label{fig:sptdim_diff_LB_CdS_v2}
\end{figure}

The return probability, obtained from the spectrum, then takes the form 
 \begin{align}
 Z(t) &\equiv \tr K(t) = \sum\limits_{v\in V} K_{v,v}(t) \nonumber\\
 & = \frac{1}{|V|} \sum\limits_{n=0}^{|V|-1} e^{-\lambda_n t / k} \, , 
 \end{align}
 where we used the decomposition in Eq.~\eqref{eq:HKernel} and the orthonormality of eigenvectors.
 
 The return probability $Z(t)$ can be nicely interpreted as a statistical \emph{partition function}, for its formal analogy with the concept in statistical physics: the diffusion time takes here the role of the inverse temperature, while the eigenvectors and their associated eigenvalues take the role of microstates and their associated energies respectively.
 
 In the case of a compact smooth manifold $\mathcal{M}$, for which the Laplace--Beltrami spectrum $\{\lambda_n\}_{n=0}^{\infty}$ is countable but unbounded, the averaged return probability density $Z(t)$ has the following asymptotic expansion for $t \rightarrow 0^+$~\cite{drumshape_rev}:
 \begin{align}\label{eq:Zmanif}
 Z(t) &= \frac{1}{\text{vol}(\mathcal{M})} \sum\limits_{n=0}^{\infty} e^{-\lambda_n t} \nonumber\\
 &= (4 \pi t)^{-\frac{\text{dim}(\mathcal{M})}{2}} \frac{1}{\text{vol}(\mathcal{M})} \Big( \sum\limits_{i=0}^{l-1}  c_i t^{\frac{i}{2}} + O(t^{\frac{l}{2}}) \Big) \, .
 \end{align}
 The return probability for unidimensional random-walks is $1/\sqrt{4 \pi t}$, so it is reasonable for a smooth manifold to locally decompose the random motion along the $\text{dim}(\mathcal{M})$ directions and get the return probability as a product of independent unidimensional return probabilities. In the case of random walks on $\mathbb{R}^d$ the return probability density equal to $Z(t)=(4 \pi t)^{-\frac{d}{2}}$, so one can infer the value of coefficients: $c_0 = \text{vol}(\mathcal{M})$ and $c_i = 0 \;\forall i \geq 1$.
 
 Corrections to the $t^{-\frac{\text{dim}(\mathcal{M})}{2}}$ behavior must be due to the geometric properties characterizing the manifold under study.
 For example, the first three coefficient have a geometrical interpretation, as discussed by McKean and Singer~\cite{heatrace_coeffs}
 \begin{align}
 c_0 &= \text{vol}(\mathcal{M}) \, ,\\
 c_1 &= -\frac{\sqrt{\pi}}{2} \text{area}(\partial \mathcal{M}) \, ,\\
 c_2 &= \frac{1}{3} \bigintsss_{\mathcal{M}} R - \frac{1}{6} \bigintsss_{\partial \mathcal{M}} J \, ,
 \end{align}
 where $\partial \mathcal{M}$ is the (possibly empty) boundary of the manifold $\mathcal{M}$, $R$ is the scalar curvature of the manifold and $J$ is the mean curvature of the boundary.
 
 We expect that similar results hold for graphs approximating manifolds, but a first difficulty can be easily detected as shown by the following argument. At a time $t$ only eigenvalues $\lambda \lesssim \frac{1}{t}$ contribute to the sum in Eq.\eqref{eq:Zmanif}, but for $t \rightarrow 0^+$ the full unbounded spectrum of the smooth manifold tends to contribute. The spectrum of a graph $G$, however, is bounded by the largest eigenvalue, so that here the expansion in Eq.~\eqref{eq:Zmanif} is not numerically reliable for times $t \lesssim (\lambda_{|V|-1})^{-1}$.
 Nevertheless one can plot the return probability as a function of time and get an estimate of the dimension $d$ by extrapolation to $\tau \rightarrow 0^+$ using the definition of what is called \emph{spectral dimension}~\cite{cdt_spectdim}:
 \begin{equation}\label{eq:sptdim_formula}
 D_S(\tau) \equiv -2 \frac{d\log Z}{d\log t}\biggr\rvert_{t=\tau}.
 \end{equation}
 
 Fig.~\ref{fig:sptdim_diff_LB_CdS_v2} shows the comparison between the estimates of spectral dimension obtained employing explicit diffusion processes (Eq.~\eqref{eq:HKernel_eq} integrated with step size $\Delta t=1$) and the spectrum of the Laplace--Beltrami matrix on graphs associated to spatial slices in $C_{dS}$ phase: we applied Eq.~\eqref{eq:sptdim_formula} using the average of the return probability $Z(t)$ computed on each slice having volume in the range $2000 - 2200$, and, for the definition via diffusion, averaging the return probability also over $200$ iterations of diffusion processes starting from randomly selected vertices in the slice. Using the definition via diffusion, at small diffusion times the return probability, and therefore also the spectral dimension, is highly fluctuating due to the short scale regularity of the tetrahedral tiling of the space (a phenomenon already discussed in Refs.~\cite{cdt_spectdim,cdt_report}); this is not present in the definition via the spectrum, where a bump is observed instead. For larger diffusion times ($\tau\gtrsim 100$) the curves obtained using both methods agree even using only the lowest $5\%$ part of the spectrum, which confirms that this regime represents indeed the large scale behavior.
 Here we observe a spectral dimension $D_S\simeq 1.5$ for the spatial slices of configurations in $C_{dS}$ phase. This fact, already observed in literature using diffusion processes~\cite{cdt_gorlich}, seems compatible also with the observations obtained from large scale scaling relations for the eigenvalues discussed in Section~\ref{subsec:scalings}.

 \end{document}